\documentclass[12pt]{article}
\usepackage{epsfig,cite,amsmath,feynarts}

\input paperdef

\def\_{\rule{.3em}{.15ex}}

\def\slash#1{\setbox0=\hbox{$#1$}#1\hskip-\wd0\dimen0=5pt\advance
       \dimen0 by-\ht0\advance\dimen0 by\dp0\lower0.5\dimen0\hbox
         to\wd0{\hss\sl/\/\hss}}

\allowdisplaybreaks

\begin{document}
\thispagestyle{empty}

\def\thefootnote{\fnsymbol{footnote}}

\begin{flushright}
CERN--PH--TH/2005--129\\
DCPT/05/102\\
IPPP/05/51\\
hep-ph/0508139\\
\end{flushright}

\vspace{1cm}

\begin{center}

{\large\sc {\bf Electroweak Precision Observables:}}
 
\vspace{0.4cm}

{\large\sc {\bf Two-Loop Yukawa Corrections of Supersymmetric
Particles}}

\vspace{1cm}

{\sc 
J.~Haestier$^{1}$%
\footnote{email: J.J.Haestier@durham.ac.uk}%
, S.~Heinemeyer$^{2}$%
\footnote{email: Sven.Heinemeyer@cern.ch}%
, D.~St\"ockinger$^{1}$%
\footnote{email: Dominik.Stockinger@durham.ac.uk}%
~and G.~Weiglein$^{1}$%
\footnote{email: Georg.Weiglein@durham.ac.uk}
}

\vspace*{1cm}

{\sl
$^1$IPPP, 
University of Durham, Durham DH1 3LE, U.K.

\vspace*{0.4cm}

$^2$CERN, TH Division, Dept. of Physics, 1211 Geneva 23, Switzerland

}

\end{center}

\vspace*{0.2cm}
\begin{abstract}
The dominant electroweak  \twol\ corrections to the 
precision observables $\MW$ and $\sweff$ are calculated in the MSSM.
They are obtained  by evaluating the two-loop Yukawa contributions of
\order{\alt^2}, \order{\alt \alb}, \order{\alb^2} to the quantity $\De\rho$.
The result, involving the contributions from Standard Model fermions,
sfermions, Higgs bosons and higgsinos,
is derived in the gauge-less limit
for arbitrary values of the lightest $\cp$-even Higgs boson mass.
A thorough discussion of the parameter relations enforced by
supersymmetry is given, and two different renormalization schemes are
applied.
Compared to the previously known result for the quark-loop contribution
we find a shift of up to $+8 \mev$ in $\MW$ and $-4 \times 10^{-5}$ in
$\sweff$. 
Detailed numerical estimates of the remaining uncertainties of $\MW$
and 
$\sweff$ from unknown higher-order contributions are obtained for
different values of the supersymmetric mass scale.

\end{abstract}

\def\thefootnote{\arabic{footnote}}
\setcounter{page}{0}
\setcounter{footnote}{0}

\newpage


\section{Introduction}

Electroweak precision observables (EWPO) like
the masses of the $W$ and $Z$ bosons, $M_{W,Z}$, or
the effective leptonic weak mixing angle, $\sweff$, are highly sensitive
probes of the quantum structure of the electroweak interactions. The
Standard Model (SM) and any extension or alternative predicts
certain relations between these observables that can be tested against
the corresponding experimental values. The experimental resolution is
better than the per-mille level~\cite{LEPEWWG}, and thus the measurements 
can be sensitive to even 
two-loop effects. Hence the EWPO are very powerful for discriminating
between different models of electroweak interactions and for deriving
indirect constraints on unknown parameters such as the masses of
the SM Higgs boson or supersymmetric particles, see \citere{PomssmRep}
for a review. A recent analysis ~\cite{ehow3} considered the EWPO,
combined with the anomalous magnetic moment of the muon and the decay
$b \to s \gamma$, in the constrained MSSM with
cold dark matter bounds. Already at the current level of 
experimental and theoretical precisions the sensitivity of the
EWPO to the scale of supersymmetry
allows to infer interesting information, pointing towards a relatively
low scale of supersymmetric particles.

The current experimental accuracies, obtained at LEP, SLC and the
Tevatron, for $\MW$, $\MZ$ and $\sweff$ are
$\de \MW=34 \mev$ (0.04\%), $\de \MZ=2.1 \mev$ (0.002\%)and 
$\de\sweff=16\times10^{-5}$ (0.07\%) \cite{LEPEWWG}. At the GigaZ
option of a linear $e^+e^-$ collider, a precision of 
$\de \MW=7 \mev$~\cite{mwgigaz,blueband} and
$\de\sweff=1.3\times10^{-5}$~\cite{swgigaz,blueband} can be achieved.

The evaluation of the SM theory predictions has reached a high level of
sophistication. The
one-loop results for the $\MW$--$\MZ$ mass relation, parametrized by the
quantity $\De r$, and for $\sweff$ are completely known. For $\De r$
also the full two-loop contributions are known~\cite{deltarSM,MWSM}, while for
$\sweff$ the calculation of all 
two-loop contributions involving a closed fermion loop 
has recently been completed~\cite{sweffSM,sweffSM2}. 
Leading universal corrections to 
the EWPO in the SM and extensions of it 
enter via a quantity called $\De\rho$. It parametrizes 
the leading universal corrections from vector boson self energies
induced by the mass splitting between fields in an isospin
doublet~\cite{rho}. Within the SM various two- and three-loop
corrections have been 
obtained~\cite{drSMgfals,drSMgfals2,drSMgf2mh0,drSMgf2mt4,ewmh2,drSMgf3mh0,drSMgf3,drSMgf3MH}. 
A contribution to $\De\rho$ induces the
following shifts in $\MW$ and $\sweff$ (with $1-\sw^2 \equiv \cw^2 =
\MW^2/\MZ^2$):
\BE
\De\MW \approx \frac{\MW}{2}\frac{\cw^2}{\cw^2 - \sw^2} \De\rho, \quad
\De\sweff \approx - \frac{\cw^2 \sw^2}{\cw^2 - \sw^2} \De\rho .
\label{precobs1}
\EE

In the minimal supersymmetric extension of the SM (MSSM)~\cite{susy}, the
theoretical evaluation of the EWPO is not as advanced as in the SM. 
In order to fully exploit the experimental precision for testing the
MSSM and deriving constraints on the supersymmetric parameters, it is
desirable to reduce the
theoretical uncertainty of the MSSM predictions to
the same level as the SM uncertainties. So
far, the one-loop contributions to $\De r$ and $\sweff$ have been
evaluated completely~\cite{dr1lA,dr1lB}. In the case of non-minimal
flavor violation the leading one-loop contributions are
known~\cite{delrhoNMFV}. At the \twol\ 
level, the leading $\oaas$ corrections to $\De\rho$ \cite{dr2lA} and
the gluonic \twol\ corrections to $\De r$ \cite{dr2lB,PomssmRep} are
known.

\begin{figure}[tb]
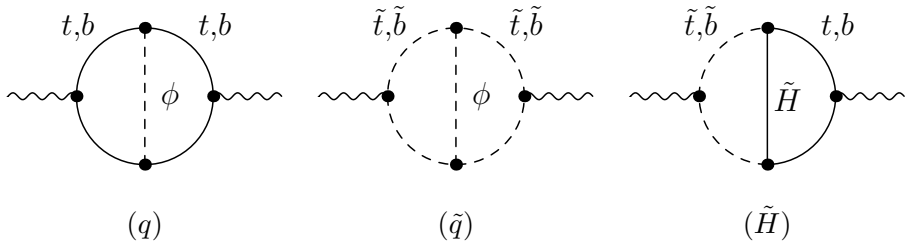

\centerline{
\unitlength=1.cm%
\begin{feynartspicture}(4,4)(1,1)
\FADiagram{$(q)$}
\FAVert(5,10){0}
\FAVert(15,10){0}
\FAProp(0,10)(5,10)(0.,){/Sine}{0}
\FAProp(15,10)(20,10)(0.,){/Sine}{0}
{
\FAProp(5,10)(15,10)(1.,){/Straight}{0}
\FAProp(15,10)(5,10)(1.,){/Straight}{0}
\FALabel(4,14)[b]{\ ${t}$,${b}$}
\FALabel(14,14)[b]{\ ${t}$,${b}$}}
\FAVert(10,15){0}
\FAVert(10,5){0}
\FAProp(10,5)(10,15)(0.,){/ScalarDash}{0}
\FALabel(11,10)[r]{$\phi$}
\end{feynartspicture}
\unitlength=1.cm%
\begin{feynartspicture}(4,4)(1,1)
\FADiagram{$(\sq)$}
\FAVert(5,10){0}
\FAVert(15,10){0}
\FAProp(0,10)(5,10)(0.,){/Sine}{0}
\FAProp(15,10)(20,10)(0.,){/Sine}{0}
{
\FAProp(5,10)(15,10)(1.,){/ScalarDash}{0}
\FAProp(15,10)(5,10)(1.,){/ScalarDash}{0}
\FALabel(4,14)[b]{\ $\tilde{t}$,$\tilde{b}$}
\FALabel(14,14)[b]{\ $\tilde{t}$,$\tilde{b}$}}
\FAVert(10,15){0}
\FAVert(10,5){0}
\FAProp(10,5)(10,15)(0.,){/ScalarDash}{0}
\FALabel(11,10)[r]{$\phi$}
\end{feynartspicture}
\unitlength=1.cm%
\begin{feynartspicture}(4,4)(1,1)
\FADiagram{$(\tilde{H})$}
\FAVert(5,10){0}
\FAVert(15,10){0}
\FAProp(0,10)(5,10)(0.,){/Sine}{0}
\FAProp(15,10)(20,10)(0.,){/Sine}{0}
{
\FAProp(5,10)(10,15)(-.4,){/ScalarDash}{0}
\FAProp(10,5)(5,10)(-.4,){/ScalarDash}{0}
\FAProp(10,15)(15,10)(-.4,){/Straight}{0}
\FAProp(10,5)(15,10)(.4,){/Straight}{0}
\FALabel(4,14)[b]{\ $\tilde{t}$,$\tilde{b}$}
\FALabel(14,14)[b]{\ ${t}$,${b}$}}
\FAVert(10,15){0}
\FAVert(10,5){0}
\FAProp(10,5)(10,15)(0.,){/Straight}{0}
\FALabel(11,10)[r]{$\tilde{H}$}
\end{feynartspicture}
}
\caption{Sample diagrams for the three classes of contributions to
  $\De\rho$ considered in this paper: $(q)$ quark loop with Higgs
  exchange, $(\sq)$ squark loop with Higgs exchange, $(\tilde{H})$
  quark/squark loop with Higgsino exchange. $\phi$ denotes Higgs and
  Goldstone boson exchange.}
\label{fig:samplediagrams}
\end{figure}

In the present paper we calculate the \twol\ MSSM-corrections to 
the EWPO that enter via $\De\rho$ at \order{\alt^2}, 
\order{\alt \alb}, \order{\alb^2}. 
These are the leading two-loop contributions involving the top and bottom
Yukawa couplings and come from three classes of diagrams with
quark/squark loop and additional Higgs or Higgsino exchange (sample
diagrams for the three classes are shown in
\reffi{fig:samplediagrams}).
These contributions are of particular interest, since they involve
corrections proportional to $\mt^4$ and bottom loop corrections
enhanced by $\tb$, the ratio of the vacuum expectation values of the two
Higgs doublets of the MSSM. Partial results have already been
presented in Ref.\ \cite{proceedings}.

As a first step, in \citere{drMSSMgf2A} the \order{\alt^2},
\order{\alt\alb}, \order{\alb^2} corrections were calculated in
the limit where the scalar quarks are heavy, corresponding to taking
into account quark/Higgs diagrams (class $(q)$) only. While this class
of corrections turned out to be well approximated by the SM
contribution (setting the Higgs-boson mass of the SM to the value of
the $\cp$-even Higgs-boson mass of the MSSM), a potentially larger 
effect can be expected from diagrams with squarks and higgsinos, 
classes $(\sq)$, $(\tilde{H})$ in \reffi{fig:samplediagrams},
which do not possess a SM counterpart.
In the
related case of similar two-loop corrections to $(g-2)_\mu$ the squark
contributions turned out to be much more important than the quark
contributions~\cite{g-2}.

For the two-loop Yukawa corrections in the SM it turned out that the 
dependence on the Higgs-boson mass is numerically important. While 
the Higgs-boson mass is a free parameter in the SM, the masses of the
$\cp$-even Higgs bosons of the MSSM are given in terms of the other
parameters of the model. In the ``gauge-less limit'' that has to be
applied in order to extract the leading two-loop Yukawa corrections, the
mass of the lighter $\cp$-even Higgs boson, $\Mh$, formally has to be
put to zero.
In \citere{drMSSMgf2A} it was observed for the calculation of the
diagrams of class $(q)$ that $\Mh$ can be set to its true
value instead of zero in a consistent way. In the
present paper we provide a detailed discussion of the gauge-less limit,
yielding an explanation of this observation. We will analyze the
Higgs-mass dependence also for the other 
classes of diagrams in \reffi{fig:samplediagrams}.

We analyze the numerical effects of the new corrections for various
scenarios in the unconstrained MSSM and for SPS benchmark
scenarios~\cite{sps}. We study two different renormalization schemes
and investigate the possible effects of unknown higher-order
corrections for $\MW$ and $\sweff$. 

The outline of the paper is as follows. First we review the
theoretical status of the EWPO, focusing on the role of $\De\rho$ and
the gauge-less limit (\refse{sec:delrho}), then we describe the
structure of the calculation (\refse{sec:details}). In
\refse{sec:renorm} the renormalization is discussed, taking into
account the implications of the gauge-less limit. 
The numerical analysis is performed in
\refse{sec:numanal}. We conclude with \refse{sec:conclusions}.

\section{Existing corrections to \boldmath{$\De\rho$} in the MSSM}
\label{sec:delrho}

The quantity $\De\rho$, 
\BE
\De\rho = \frac{\Si_Z(0)}{\MZ^2} - \frac{\Si_W(0)}{\MW^2} ,
\label{delrho}
\EE
 parametrizes the leading universal corrections to the electroweak
 precision observables induced by
the mass splitting between fields in an isospin doublet~\cite{rho}.
$\Si_{Z,W}(0)$ denote the transverse parts of the unrenormalized $Z$
 and $W$ boson self-energies at zero momentum transfer, respectively.
The corresponding shifts in the predictions of $\MW$ and $\sweff$ have
been given in \refeq{precobs1}.

A detailed overview of the existing corrections to the EWPO
within the SM and the MSSM can be found in \citere{PomssmRep}. 
Here we briefly review the existing corrections 
to $\De\rho$ in the MSSM and the corresponding SM contributions.
Since $\De\rho$ as defined in \refeq{delrho} is not a physical
observable but contains only a part of the quantum corrections to the
EWPO, some care has to be taken in its evaluation. 
Only contributions that form a complete subset of leading corrections
to the EWPO can be consistently incorporated into $\De\rho$.
While the fermion loop contributions to $\De\rho$ form a meaningful
subset of the full one-loop corrections to the EWPO (see below), a naive
evaluation of \refeq{delrho} for the
bosonic one-loop contributions of the SM
would result in an expression that is neither
UV-finite nor gauge-parameter independent (see the discussion in
\citeres{PomssmRep,Denner:1995jd}). A consistent inclusion of the
bosonic corrections therefore makes it necessary to go
beyond the $\De\rho$
approximation. We will introduce the so-called ``gauge-less limit'' in 
\refse{sec:ewgaugeless} and show that at
\order{\alt^2}, \order{\alt\alb}, \order{\alb^2} 
the contributions to $\De\rho$ coincide with the
complete contributions to the EWPO.


\subsection{One-loop and higher-order QCD corrections to $\De\rho$}

The dominant one-loop contribution to $\De\rho$ within the SM 
arises from the top/bottom doublet~\cite{rho}. It is given by
\BE
\De\rho^{\SM}_{\rm 1-loop} = \frac{3\,\gf}{8\,\wz\,\pi^2} \; 
F_0(\mt^2, \mb^2) ,
\label{eq:delrhoSM1l}
\EE
where
\BE
F_0(x,y) = x + y - \frac{2\,x\,y}{x - y} \log \frac{x}{y} .
\EE
$F_0$ has the properties $F_0(m_1^2, m_2^2) = F_0(m_2^2, m_1^2)$, 
$F_0(m^2, m^2) = 0$, $F_0(m^2, 0) = m^2$. 
One therefore obtains $F_0(\mt^2, \mb^2) \approx \mt^2$, giving rise to
the well-known quadratic dependence of the one-loop corrections to the 
EWPO on the top-quark mass.

Within the MSSM the dominant correction from supersymmetric particles at the
\onel\ level arises from 
the scalar top and bottom contribution to \refeq{delrho}. For 
$\mb \neq 0$ it is given by
\begin{align}
\De\rho_{\rm 1-loop}^\SU = \frac{3\,\gf}{8\, \wz\, \pi^2} 
                  \Big[& -\sinQtt\cosQtt F_0(\mste^2, \mstz^2)
                       -\sinQtb\cosQtb F_0(\msbe^2, \msbz^2) \non\\
                       &+\cosQtt\cosQtb F_0(\mste^2, \msbe^2)
                       +\cosQtt\sinQtb F_0(\mste^2, \msbz^2) \non\\
                       &+\sinQtt\cosQtb F_0(\mstz^2, \msbe^2)
                       +\sinQtt\sinQtb F_0(\mstz^2, \msbz^2) ~\Big] .
\label{delrhoMSSM1l}
\end{align}
Here $\msti, \msbi (i = 1,2)$ denote the stop and sbottom masses,
whereas $\tst, \tsb$ are the mixing angles in the stop and in the
sbottom sector, see also \refse{subsec:squark}.

\smallskip
The dominant \twol\ correction within the SM of 
\order{\al\als} is given by~\cite{drSMgfals}
\BE
\De\rho^{\SM, \al\als}_{\rm 2-loop} = - \De\rho^{\SM}_{\rm 1-loop} \frac{2}{3} \frac{\als}{\pi}
                           \KL 1 + \pi^2/3 \KR .
\EE
It screens the \onel\ result by approximately 10\%. 

The corresponding corrections of \order{\al\als} in the MSSM have been
evaluated in \citere{dr2lA}. 
Contrary to the SM case, these corrections can enter with the same sign as 
the one-loop result, therefore enhancing the sensitivity to the squark effects.


\subsection{Electroweak two-loop corrections to $\De\rho$: the
gauge-less limit}
\label{sec:ewgaugeless}

The Yukawa contributions of \order{\al_f^2} form a set of leading two-loop 
contributions entering the EWPO only via $\De\rho$, where 
$\al_f \equiv y_f^2/(4 \pi)$, and $y_f$ is the Yukawa coupling of 
fermion $f$. 
For the top and bottom quarks the Yukawa couplings read
\BE
y_t = \frac{\wz \, \mt}{v \, \Sb}, \quad
y_b = \frac{\wz \, \mb}{v \, \Cb} ~,
\label{ytyb}
\EE
where $v \equiv \sqrt{v_1^2 + v_2^2}$. 
In the SM another subset of leading
electroweak two-loop corrections to $\De\rho$ is given by the 
corrections for large Higgs-boson masses of
\order{\gf^2\MH^2\MW^2}~\cite{ewmh2}. We will focus on the
Yukawa corrections in the following.

In order to evaluate 
the leading Yukawa contributions of \order{\al_f^2}
the gauge-less limit has to be applied. 
It consists of neglecting the electroweak gauge couplings
$g_{1,2}\to0$ and thus also $\MW^2 = g_2^2 v^2/2\to0$ and
$\MZ^2 = (g_1^2 + g_2^2) v^2/2\to0$, while keeping the ratio 
$\cw = \MW/\MZ$ and the vacuum expectation value $v$
fixed. Accordingly, $\Sigma_{Z,W}$ in \refeq{delrho} need to be evaluated
at \order{g_{1,2}^2} in order to obtain a finite contribution of
\order{g_{1,2}^0} to $\De\rho$ in the gauge-less limit.
In this limit only diagrams with fermions and scalars contribute to
$\De\rho$, while no gauge bosons appear in the loop diagrams.

At the one-loop level the only non-vanishing contributions to $\De\rho$
in the gauge-less limit of the MSSM are the 
fermion-loop and sfermion-loop contributions as given in 
\refeqs{eq:delrhoSM1l}, (\ref{delrhoMSSM1l}). While the Higgs sector 
of a general two-Higgs-doublet model yields a contribution to $\De\rho$
in the gauge-less limit, the contribution vanishes once the symmetry
relations of the MSSM are imposed (see the discussion in \refse{HiggsSecRen}
below).

At the two-loop level the gauge-less limit results in the desired Yukawa
contributions of \order{\alpha_f^2}. For the quarks and squarks of the 
third generation this yields in particular terms of \order{\mt^4/v^4}
and \order{\mb^4\tan^2\be/v^4}. It is easy
to see that no other contribution to $\MW$ and $\sweff$ besides
the gauge-less limit of $\De\rho$ yields terms of
this order.

\smallskip 
In the SM the two-loop result for $\De\rho$ in the gauge-less limit was
first obtained for the special case $\MHSM = 0$~\cite{drSMgf2mh0},
\BE
\De\rho^{\SM,\al_t^2}_{{\rm 2-loop} |\MH = 0} =
 3\,\frac{\gf^2}{128 \pi^4} \, \mt^4\, \KL 19 - 2 \pi^2 \KR~.
\label{SMMH0}
\EE
This result was then extended to the case of arbitrary values of
$\MHSM$~\cite{drSMgf2mt4}. The corresponding result is given by
\BEA
\label{drSMgf2}
\De\rho^{\SM,\al_t^2}_{\rm 2-loop}(\MHSM) &=&
 3\,\frac{\gf^2}{128 \pi^4} \, \mt^4\, 
\biggl\{   -3 \frac{x^2 (10 - 6 x^2 + x^4)}{x^2 - 4}
    w \log^2\KL\frac{1-w}{2}\KR \\
&&\mbox{} + \frac{x^2 (x^2 - 4)}{2}
    w \log\KL\frac{1 - w}{1 + w}\KR  \non \\
&&\mbox{} - \KL 6 + 6 x^2 - \frac{x^4}{2} \KR \log\KL x^2 \KR 
    + 3 \frac{x^2 (10 - 6 x^2 + x^4)}{2 (x^2 -4)} 
      w \log^2\KL x^2 \KR \non \\
&&\mbox{} + 25 - 4 x^2 
 +  \pi^2 \frac{ + 8 - 6 x^2  + x^4 
              -10 w x^4 + 6 w x^6 - w x^8 }{2 x^2 (x^2 - 4)} \non \\
&&\mbox{} + 6 w x^2 \frac{10 - 6 x^2 + x^4}{x^2 - 4} 
    {\rm Li}_2\KL\frac{1 - w}{2}\KR \non \\
&&\mbox{} - \frac{3 (x^2 - 1)^2 (x^2 - 2)}{x^2}
    {\rm Li}_2\KL 1-x^2\KR \biggr\} , \non
\EEA
where $x = \MHSM/\mt$, and $w = \sqrt{1 - 4/x^2}$. 
The effect of going beyond the approximation $\MHSM = 0$ turned out to
be numerically very significant. While the numerical value of the result
in \refeq{SMMH0} is rather small due to the accidental cancellation of
the two terms in the last factor of \refeq{SMMH0}, the result is about
an order of magnitude larger
for values of $\MHSM$ in the experimentally preferred region.
As a consequence, the result for the \order{\al_t^2} corrections to 
$\De\rho$ with arbitrary
Higgs-boson mass as given in \refeq{drSMgf2} provides a much better
approximation of the full electroweak two-loop corrections to the
EWPO~\cite{deltarSM,MWSM,sweffSM,sweffSM2} than the limiting case 
where $\MHSM = 0$, \refeq{SMMH0}. 
As an example, for $\MHSM=120\gev$ the resulting shifts in $\MW$ and
$\sweff$ are $-10\mev$ and $+5\times10^{-5}$, respectively.

\smallskip 
Within the MSSM also the contributions involving the bottom Yukawa
coupling can be relevant at large $\tb$. The corresponding
contributions of \order{\alt^2}, \order{\alt \alb}, and \order{\alb^2}
to $\De\rho$ have been obtained in \citere{drMSSMgf2A}
in the limit of heavy scalar quarks. In this limit only the top and
bottom quarks and the Higgs bosons (and Goldstone bosons) 
of the MSSM appear in the loops. The results turned out to be
numerically relevant, leading to shifts in $\MW$ and $\sweff$ of up to 
$12 \mev$ and $6 \times 10^{-5}$, respectively. Since in the
gauge-less limit the couplings of the light $\cp$-even Higgs boson of
the MSSM to fermions become SM-like, the \order{\alt^2} correction in
the MSSM can be well approximated by the corresponding correction in the
SM, as given in \refeq{drSMgf2}. Potentially larger effects compared to
the SM case can be
expected from the contribution of supersymmetric particles (with not too
heavy masses), since these corrections do not have a SM counterpart.


\section{The \boldmath{\order{\alt^2}, \order{\alt \alb},
            \order{\alb^2}}  contributions to \boldmath{$\De\rho$}}
\label{sec:details}

The purpose of the present paper is to perform a complete calculation
of the \order{\alt^2}, \order{\alt \alb}, and \order{\alb^2}
contributions to $\De\rho$ in the MSSM, including the contributions of
supersymmetric 
particles. This means that all diagrams have to be evaluated (applying
the gauge-less limit) that contain top and bottom quarks, their scalar
superpartners stop and sbottom, and Higgs bosons or higgsinos.

\begin{figure}[htb!]
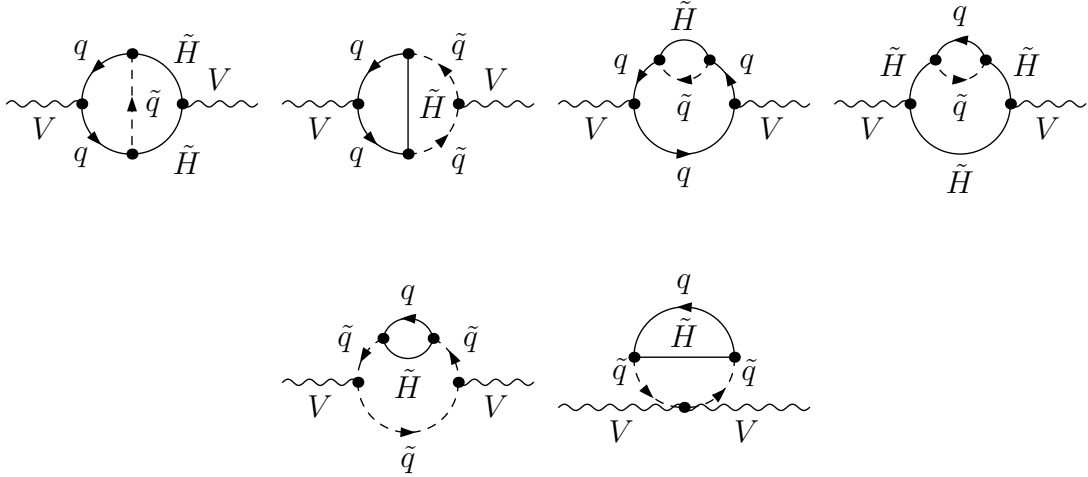

\vspace{2em}
\begin{center}
\unitlength=1bp%

\begin{feynartspicture}(432,104)(4,1)

\FADiagram{}
\FAProp(0.,10.)(6.,10.)(0.,){/Sine}{0}
\FALabel(3.,8.93)[t]{$V$}
\FAProp(20.,10.)(14.,10.)(0.,){/Sine}{0}
\FALabel(17.,11.07)[b]{$V$}
\FAProp(10.,6.)(6.,10.)(-0.434885,){/Straight}{-1}
\FALabel(6.51421,6.51421)[tr]{$q$}
\FAProp(10.,6.)(14.,10.)(0.412689,){/Straight}{0}
\FALabel(13.2646,6.73538)[tl]{$\tilde{H}$}
\FAProp(10.,14.)(10.,6.)(0.,){/ScalarDash}{-1}
\FALabel(11.07,10.)[l]{$\sq$}
\FAProp(10.,14.)(6.,10.)(0.425735,){/Straight}{1}
\FALabel(6.53252,13.4675)[br]{$q$}
\FAProp(10.,14.)(14.,10.)(-0.412689,){/Straight}{0}
\FALabel(13.2646,13.2646)[bl]{$\tilde{H}$}
\FAVert(10.,14.){0}
\FAVert(10.,6.){0}
\FAVert(6.,10.){0}
\FAVert(14.,10.){0}

\FADiagram{}
\FAProp(0.,10.)(6.,10.)(0.,){/Sine}{0}
\FALabel(3.,8.93)[t]{$V$}
\FAProp(20.,10.)(14.,10.)(0.,){/Sine}{0}
\FALabel(17.,11.07)[b]{$V$}
\FAProp(10.,6.)(6.,10.)(-0.434885,){/Straight}{-1}
\FALabel(6.51421,6.51421)[tr]{$q$}
\FAProp(10.,6.)(14.,10.)(0.412689,){/ScalarDash}{1}
\FALabel(13.4414,6.55861)[tl]{$\sq$}
\FAProp(10.,14.)(10.,6.)(0.,){/Straight}{0}
\FALabel(10.82,10.)[l]{$\tilde{H}$}
\FAProp(10.,14.)(6.,10.)(0.425735,){/Straight}{1}
\FALabel(6.53252,13.4675)[br]{$q$}
\FAProp(10.,14.)(14.,10.)(-0.412689,){/ScalarDash}{-1}
\FALabel(13.4414,13.4414)[bl]{$\sq$}
\FAVert(10.,14.){0}
\FAVert(10.,6.){0}
\FAVert(6.,10.){0}
\FAVert(14.,10.){0}

\FADiagram{}
\FAProp(0.,10.)(6.,10.)(0.,){/Sine}{0}
\FALabel(3.,8.93)[t]{$V$}
\FAProp(20.,10.)(14.,10.)(0.,){/Sine}{0}
\FALabel(17.,8.93)[t]{$V$}
\FAProp(6.,10.)(14.,10.)(1.,){/Straight}{1}
\FALabel(10.,4.93)[t]{$q$}
\FAProp(8.,13.5)(6.,10.)(0.315846,){/Straight}{1}
\FALabel(5.58149,12.3549)[br]{$q$}
\FAProp(12.,13.5)(8.,13.5)(0.8,){/Straight}{0}
\FALabel(10.,15.92)[b]{$\tilde{H}$}
\FAProp(12.,13.5)(8.,13.5)(-0.8,){/ScalarDash}{1}
\FALabel(10.,10.83)[t]{$\sq$}
\FAProp(12.,13.5)(14.,10.)(-0.310119,){/Straight}{-1}
\FALabel(14.4085,12.3491)[bl]{$q$}
\FAVert(12.,13.5){0}
\FAVert(8.,13.5){0}
\FAVert(6.,10.){0}
\FAVert(14.,10.){0}

\FADiagram{}
\FAProp(0.,10.)(6.,10.)(0.,){/Sine}{0}
\FALabel(3.,8.93)[t]{$V$}
\FAProp(20.,10.)(14.,10.)(0.,){/Sine}{0}
\FALabel(17.,8.93)[t]{$V$}
\FAProp(6.,10.)(14.,10.)(1.,){/Straight}{0}
\FALabel(10.,5.18)[t]{$\tilde{H}$}
\FAProp(8.,13.5)(6.,10.)(0.315846,){/Straight}{0}
\FALabel(5.79855,12.2308)[br]{$\tilde{H}$}
\FAProp(12.,13.5)(8.,13.5)(0.8,){/Straight}{1}
\FALabel(10.,16.17)[b]{$q$}
\FAProp(12.,13.5)(8.,13.5)(-0.8,){/ScalarDash}{-1}
\FALabel(10.,10.83)[t]{$\sq$}
\FAProp(12.,13.5)(14.,10.)(-0.310119,){/Straight}{0}
\FALabel(14.1914,12.2251)[bl]{$\tilde{H}$}
\FAVert(12.,13.5){0}
\FAVert(8.,13.5){0}
\FAVert(6.,10.){0}
\FAVert(14.,10.){0}
\end{feynartspicture}

\begin{feynartspicture}(432,104)(4,1)

\FADiagram{}

\FADiagram{}
\FAProp(0.,10.)(6.,10.)(0.,){/Sine}{0}
\FALabel(3.,8.93)[t]{$V$}
\FAProp(20.,10.)(14.,10.)(0.,){/Sine}{0}
\FALabel(17.,8.93)[t]{$V$}
\FAProp(6.,10.)(14.,10.)(1.,){/ScalarDash}{1}
\FALabel(10.,4.93)[t]{$\sq$}
\FAProp(8.,13.5)(6.,10.)(0.315846,){/ScalarDash}{1}
\FALabel(5.58149,12.3549)[br]{$\sq$}
\FAProp(12.,13.5)(8.,13.5)(0.8,){/Straight}{1}
\FALabel(10.,16.17)[b]{$q$}
\FAProp(12.,13.5)(8.,13.5)(-0.8,){/Straight}{0}
\FALabel(10.,11.08)[t]{$\tilde{H}$}
\FAProp(12.,13.5)(14.,10.)(-0.310119,){/ScalarDash}{-1}
\FALabel(14.4085,12.3491)[bl]{$\sq$}
\FAVert(12.,13.5){0}
\FAVert(8.,13.5){0}
\FAVert(6.,10.){0}
\FAVert(14.,10.){0}

\FADiagram{}
\FAProp(0.,8.)(10.,8.)(0.,){/Sine}{0}
\FALabel(5.,6.93)[t]{$V$}
\FAProp(20.,8.)(10.,8.)(0.,){/Sine}{0}
\FALabel(15.,6.93)[t]{$V$}
\FAProp(6.,12.)(14.,12.)(-1.,){/Straight}{-1}
\FALabel(10.,17.07)[b]{$q$}
\FAProp(6.,12.)(14.,12.)(0.,){/Straight}{0}
\FALabel(10.,12.82)[b]{$\tilde{H}$}
\FAProp(6.,12.)(10.,8.)(0.431748,){/ScalarDash}{1}
\FALabel(4.25252,9.75252)[bl]{$\sq$}
\FAProp(14.,12.)(10.,8.)(-0.450694,){/ScalarDash}{-1}
\FALabel(15.7854,9.71463)[br]{$\sq$}
\FAVert(6.,12.){0}
\FAVert(14.,12.){0}
\FAVert(10.,8.){0}

\FADiagram{}

\end{feynartspicture}

\end{center}
\caption[]{
Generic Feynman diagrams of class $(\tilde{H})$. 
$V$ denotes either $W$ or $Z$, $\sq$ is either a $\Stop$ or a $\Sbot$,
while $q$ is a $t$ or a $b$, and $\tilde H$ denotes a higgsino
(neutral or charged).
}
\label{fig:generichiggsino}
\end{figure}

The contributions to $\De\rho$ 
at \order{\alt^2}, \order{\alt \alb}, \order{\alb^2} 
can be grouped into 
three classes (see \reffi{fig:samplediagrams}):
\begin{itemize}
\item[($q$)]  
diagrams involving $t/b$ quarks and Higgs bosons (see also
\citere{drMSSMgf2A}),
\item[($\sq$)] 
diagrams with $\Stop/\Sbot$ squarks and Higgs bosons (see
\reffi{fig:genericsquark} for generic diagrams),
\item[($\tilde H$)] 
diagrams with higgsinos (containing also quarks and squarks) (see
\reffi{fig:generichiggsino} for generic diagrams).
\end{itemize}

The generic diagrams  shown in 
\reffis{fig:genericsquark}, \ref{fig:generichiggsino} 
have to be evaluated for the $Z$ 
boson and the $W$ boson self-energy.

\begin{figure}[htb!]
\begin{center}
\unitlength=1bp%

\begin{feynartspicture}(432,104)(4,1)

\FADiagram{}
\FAProp(0.,10.)(6.,10.)(0.,){/Sine}{0}
\FALabel(3.,8.93)[t]{$V$}
\FAProp(20.,10.)(14.,10.)(0.,){/Sine}{0}
\FALabel(17.,8.93)[t]{$V$}
\FAProp(6.,10.)(14.,10.)(1.,){/ScalarDash}{0}
\FALabel(10.,5.18)[t]{$\chi$}
\FAProp(8.,13.5)(6.,10.)(0.315846,){/ScalarDash}{0}
\FALabel(5.79855,12.2308)[br]{$\phi$}
\FAProp(12.,13.5)(8.,13.5)(0.8,){/ScalarDash}{-1}
\FALabel(10.,16.17)[b]{$\sq$}
\FAProp(12.,13.5)(8.,13.5)(-0.8,){/ScalarDash}{1}
\FALabel(10.,10.83)[t]{$\sq$}
\FAProp(12.,13.5)(14.,10.)(-0.310119,){/ScalarDash}{0}
\FALabel(14.1914,12.2251)[bl]{$\phi$}
\FAVert(12.,13.5){0}
\FAVert(8.,13.5){0}
\FAVert(6.,10.){0}
\FAVert(14.,10.){0}

\FADiagram{}
\FAProp(0.,10.)(6.,10.)(0.,){/Sine}{0}
\FALabel(3.,8.93)[t]{$V$}
\FAProp(20.,10.)(14.,10.)(0.,){/Sine}{0}
\FALabel(17.,8.93)[t]{$V$}
\FAProp(6.,10.)(14.,10.)(1.,){/ScalarDash}{1}
\FALabel(10.,4.93)[t]{$\sq$}
\FAProp(8.,13.5)(6.,10.)(0.315846,){/ScalarDash}{1}
\FALabel(5.58149,12.3549)[br]{$\sq$}
\FAProp(12.,13.5)(8.,13.5)(0.8,){/ScalarDash}{0}
\FALabel(10.,15.92)[b]{$\phi$}
\FAProp(12.,13.5)(8.,13.5)(-0.8,){/ScalarDash}{1}
\FALabel(10.,10.83)[t]{$\sq$}
\FAProp(12.,13.5)(14.,10.)(-0.310119,){/ScalarDash}{-1}
\FALabel(14.4085,12.3491)[bl]{$\sq$}
\FAVert(12.,13.5){0}
\FAVert(8.,13.5){0}
\FAVert(6.,10.){0}
\FAVert(14.,10.){0}

\FADiagram{}
\FAProp(0.,10.)(6.,10.)(0.,){/Sine}{0}
\FALabel(3.,8.93)[t]{$V$}
\FAProp(20.,10.)(14.,10.)(0.,){/Sine}{0}
\FALabel(17.,11.07)[b]{$V$}
\FAProp(10.,6.)(6.,10.)(-0.434885,){/ScalarDash}{0}
\FALabel(6.69099,6.69099)[tr]{$\phi$}
\FAProp(10.,6.)(14.,10.)(0.412689,){/ScalarDash}{1}
\FALabel(13.4414,6.55861)[tl]{$\sq$}
\FAProp(10.,14.)(10.,6.)(0.,){/ScalarDash}{1}
\FALabel(11.07,10.)[l]{$\sq$}
\FAProp(10.,14.)(6.,10.)(0.425735,){/ScalarDash}{0}
\FALabel(6.70929,13.2907)[br]{$\chi$}
\FAProp(10.,14.)(14.,10.)(-0.412689,){/ScalarDash}{-1}
\FALabel(13.4414,13.4414)[bl]{$\sq$}
\FAVert(10.,14.){0}
\FAVert(10.,6.){0}
\FAVert(6.,10.){0}
\FAVert(14.,10.){0}

\FADiagram{}
\FAProp(0.,10.)(6.,10.)(0.,){/Sine}{0}
\FALabel(3.,8.93)[t]{$V$}
\FAProp(20.,10.)(14.,10.)(0.,){/Sine}{0}
\FALabel(17.,11.07)[b]{$V$}
\FAProp(10.,6.)(6.,10.)(-0.434885,){/ScalarDash}{-1}
\FALabel(6.51421,6.51421)[tr]{$\sq$}
\FAProp(10.,6.)(14.,10.)(0.412689,){/ScalarDash}{1}
\FALabel(13.4414,6.55861)[tl]{$\sq$}
\FAProp(10.,14.)(10.,6.)(0.,){/ScalarDash}{0}
\FALabel(10.82,10.)[l]{$\phi$}
\FAProp(10.,14.)(6.,10.)(0.425735,){/ScalarDash}{1}
\FALabel(6.53252,13.4675)[br]{$\sq$}
\FAProp(10.,14.)(14.,10.)(-0.412689,){/ScalarDash}{-1}
\FALabel(13.4414,13.4414)[bl]{$\sq$}
\FAVert(10.,14.){0}
\FAVert(10.,6.){0}
\FAVert(6.,10.){0}
\FAVert(14.,10.){0}
\end{feynartspicture}

\begin{feynartspicture}(432,104)(4,1)

\FADiagram{}
\FAProp(0.,8.)(10.,8.)(0.,){/Sine}{0}
\FALabel(5.,6.93)[t]{$V$}
\FAProp(20.,8.)(10.,8.)(0.,){/Sine}{0}
\FALabel(15.,6.93)[t]{$V$}
\FAProp(6.,12.)(14.,12.)(-1.,){/ScalarDash}{0}
\FALabel(10.,16.82)[b]{$\phi$}
\FAProp(6.,12.)(14.,12.)(0.,){/ScalarDash}{-1}
\FALabel(10.,13.07)[b]{$\sq$}
\FAProp(6.,12.)(10.,8.)(0.431748,){/ScalarDash}{1}
\FALabel(4.25252,9.75252)[bl]{$\sq$}
\FAProp(14.,12.)(10.,8.)(-0.450694,){/ScalarDash}{-1}
\FALabel(15.7854,9.71463)[br]{$\sq$}
\FAVert(6.,12.){0}
\FAVert(14.,12.){0}
\FAVert(10.,8.){0}

\FADiagram{}
\FAProp(0.,9.)(6.,9.)(0.,){/Sine}{0}
\FALabel(3.,7.93)[t]{$V$}
\FAProp(20.,9.)(14.,9.)(0.,){/Sine}{0}
\FALabel(17.,7.93)[t]{$V$}
\FAProp(6.,9.)(14.,9.)(1.,){/ScalarDash}{-1}
\FALabel(10.,3.93)[t]{$\sq$}
\FAProp(6.,9.)(10.,13.)(-0.440636,){/ScalarDash}{1}
\FALabel(6.50271,12.4973)[br]{$\sq$}
\FAProp(14.,9.)(10.,13.)(0.425735,){/ScalarDash}{-1}
\FALabel(13.4675,12.4675)[bl]{$\sq$}
\FAProp(10.,13.)(10.,13.)(10.,17.){/ScalarDash}{-1}
\FALabel(10.,18.07)[b]{$\sq$}
\FAVert(6.,9.){0}
\FAVert(14.,9.){0}
\FAVert(10.,13.){0}

\FADiagram{}
\FAProp(0.,9.)(6.,9.)(0.,){/Sine}{0}
\FALabel(3.,7.93)[t]{$V$}
\FAProp(20.,9.)(14.,9.)(0.,){/Sine}{0}
\FALabel(17.,7.93)[t]{$V$}
\FAProp(6.,9.)(14.,9.)(1.,){/ScalarDash}{0}
\FALabel(10.,4.18)[t]{$\chi$}
\FAProp(6.,9.)(10.,13.)(-0.440636,){/ScalarDash}{0}
\FALabel(6.67949,12.3205)[br]{$\phi$}
\FAProp(14.,9.)(10.,13.)(0.425735,){/ScalarDash}{0}
\FALabel(13.2907,12.2907)[bl]{$\phi$}
\FAProp(10.,13.)(10.,13.)(10.,17.){/ScalarDash}{-1}
\FALabel(10.,18.07)[b]{$\sq$}
\FAVert(6.,9.){0}
\FAVert(14.,9.){0}
\FAVert(10.,13.){0}

\FADiagram{}
\FAProp(0.,9.)(6.,9.)(0.,){/Sine}{0}
\FALabel(3.,7.93)[t]{$V$}
\FAProp(20.,9.)(14.,9.)(0.,){/Sine}{0}
\FALabel(17.,7.93)[t]{$V$}
\FAProp(6.,9.)(14.,9.)(1.,){/ScalarDash}{-1}
\FALabel(10.,3.93)[t]{$\sq$}
\FAProp(6.,9.)(10.,13.)(-0.440636,){/ScalarDash}{1}
\FALabel(6.50271,12.4973)[br]{$\sq$}
\FAProp(14.,9.)(10.,13.)(0.425735,){/ScalarDash}{-1}
\FALabel(13.4675,12.4675)[bl]{$\sq$}
\FAProp(10.,13.)(10.,13.)(10.,17.){/ScalarDash}{0}
\FALabel(10.,17.82)[b]{$\phi$}
\FAVert(6.,9.){0}
\FAVert(14.,9.){0}
\FAVert(10.,13.){0}
\end{feynartspicture}

\begin{feynartspicture}(432,104)(4,1)

\FADiagram{}
\FAProp(0.,8.)(10.,8.)(0.,){/Sine}{0}
\FALabel(5.,6.93)[t]{$V$}
\FAProp(20.,8.)(10.,8.)(0.,){/Sine}{0}
\FALabel(15.,6.93)[t]{$V$}
\FAProp(6.,12.)(14.,12.)(-1.,){/ScalarDash}{-1}
\FALabel(10.,17.07)[b]{$\sq$}
\FAProp(6.,12.)(14.,12.)(0.,){/ScalarDash}{1}
\FALabel(10.,13.07)[b]{$\sq$}
\FAProp(6.,12.)(10.,8.)(0.431748,){/ScalarDash}{0}
\FALabel(4.07574,9.57574)[bl]{$\phi$}
\FAProp(14.,12.)(10.,8.)(-0.450694,){/ScalarDash}{0}
\FALabel(15.9621,9.53785)[br]{$\phi$}
\FAVert(6.,12.){0}
\FAVert(14.,12.){0}
\FAVert(10.,8.){0}

\FADiagram{}
\FAProp(0.,7.)(10.,7.)(0.,){/Sine}{0}
\FALabel(5.,5.93)[t]{$V$}
\FAProp(20.,7.)(10.,7.)(0.,){/Sine}{0}
\FALabel(15.,5.93)[t]{$V$}
\FAProp(10.,7.)(10.,12.)(-1.,){/ScalarDash}{-1}
\FALabel(6.43,9.5)[r]{$\sq$}
\FAProp(10.,7.)(10.,12.)(1.,){/ScalarDash}{1}
\FALabel(13.57,9.5)[l]{$\sq$}
\FAProp(10.,12.)(10.,12.)(10.,17.){/ScalarDash}{-1}
\FALabel(10.,18.07)[b]{$\sq$}
\FAVert(10.,7.){0}
\FAVert(10.,12.){0}

\FADiagram{}
\FAProp(0.,7.)(10.,7.)(0.,){/Sine}{0}
\FALabel(5.,5.93)[t]{$V$}
\FAProp(20.,7.)(10.,7.)(0.,){/Sine}{0}
\FALabel(15.,5.93)[t]{$V$}
\FAProp(10.,7.)(10.,12.)(-1.,){/ScalarDash}{0}
\FALabel(6.68,9.5)[r]{$\phi$}
\FAProp(10.,7.)(10.,12.)(1.,){/ScalarDash}{0}
\FALabel(13.32,9.5)[l]{$\phi$}
\FAProp(10.,12.)(10.,12.)(10.,17.){/ScalarDash}{-1}
\FALabel(10.,18.07)[b]{$\sq$}
\FAVert(10.,7.){0}
\FAVert(10.,12.){0}

\FADiagram{}
\FAProp(0.,7.)(10.,7.)(0.,){/Sine}{0}
\FALabel(5.,5.93)[t]{$V$}
\FAProp(20.,7.)(10.,7.)(0.,){/Sine}{0}
\FALabel(15.,5.93)[t]{$V$}
\FAProp(10.,7.)(10.,12.)(-1.,){/ScalarDash}{-1}
\FALabel(6.43,9.5)[r]{$\sq$}
\FAProp(10.,7.)(10.,12.)(1.,){/ScalarDash}{1}
\FALabel(13.57,9.5)[l]{$\sq$}
\FAProp(10.,12.)(10.,12.)(10.,17.){/ScalarDash}{0}
\FALabel(10.,17.82)[b]{$\phi$}
\FAVert(10.,7.){0}
\FAVert(10.,12.){0}
\end{feynartspicture}

\end{center}

\caption[]{
Generic Feynman diagrams of class $(\sq)$. 
$V$ denotes either $W$ or $Z$, $\sq$ is either a $\Stop$ or a $\Sbot$,
and $\phi,\chi$ denote Higgs and Goldstone bosons.
}
\label{fig:genericsquark}
\end{figure}

\begin{figure}[htb!]
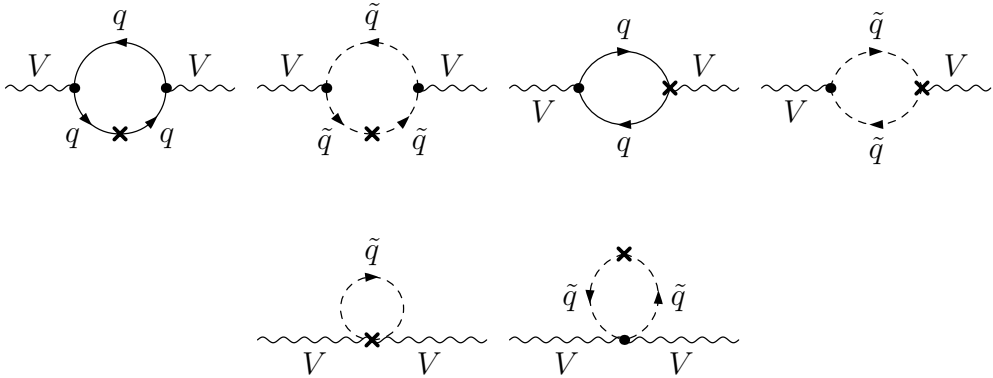

\begin{center}

\unitlength=1bp%

\begin{feynartspicture}(432,190)(4,2)

\FADiagram{}
\FAProp(0.,10.)(6.,10.)(0.,){/Sine}{0}
\FALabel(3.,11.07)[b]{$V$}
\FAProp(20.,10.)(14.,10.)(0.,){/Sine}{0}
\FALabel(17.,11.07)[b]{$V$}
\FAProp(10.,6.)(6.,10.)(-0.431295,){/Straight}{-1}
\FALabel(6.52139,6.52139)[tr]{$q$}
\FAProp(10.,6.)(14.,10.)(0.431295,){/Straight}{1}
\FALabel(13.4786,6.52139)[tl]{$q$}
\FAProp(6.,10.)(14.,10.)(-1.,){/Straight}{-1}
\FALabel(10.,15.07)[b]{$q$}
\FAVert(6.,10.){0}
\FAVert(14.,10.){0}
\FAVert(10.,6.){1}

\FADiagram{}
\FAProp(0.,10.)(6.,10.)(0.,){/Sine}{0}
\FALabel(3.,11.07)[b]{$V$}
\FAProp(20.,10.)(14.,10.)(0.,){/Sine}{0}
\FALabel(17.,11.07)[b]{$V$}
\FAProp(10.,6.)(6.,10.)(-0.431295,){/ScalarDash}{-1}
\FALabel(6.52139,6.52139)[tr]{$\sq$}
\FAProp(10.,6.)(14.,10.)(0.431295,){/ScalarDash}{1}
\FALabel(13.4786,6.52139)[tl]{$\sq$}
\FAProp(6.,10.)(14.,10.)(-1.,){/ScalarDash}{-1}
\FALabel(10.,15.07)[b]{$\sq$}
\FAVert(6.,10.){0}
\FAVert(14.,10.){0}
\FAVert(10.,6.){1}

\FADiagram{}
\FAProp(0.,10.)(6.,10.)(0.,){/Sine}{0}
\FALabel(3.,8.93)[t]{$V$}
\FAProp(20.,10.)(14.,10.)(0.,){/Sine}{0}
\FALabel(17.,11.07)[b]{$V$}
\FAProp(6.,10.)(14.,10.)(0.8,){/Straight}{-1}
\FALabel(10.,5.73)[t]{$q$}
\FAProp(6.,10.)(14.,10.)(-0.8,){/Straight}{1}
\FALabel(10.,14.27)[b]{$q$}
\FAVert(6.,10.){0}
\FAVert(14.,10.){1}

\FADiagram{}
\FAProp(0.,10.)(6.,10.)(0.,){/Sine}{0}
\FALabel(3.,8.93)[t]{$V$}
\FAProp(20.,10.)(14.,10.)(0.,){/Sine}{0}
\FALabel(17.,11.07)[b]{$V$}
\FAProp(6.,10.)(14.,10.)(0.8,){/ScalarDash}{-1}
\FALabel(10.,5.73)[t]{$\sq$}
\FAProp(6.,10.)(14.,10.)(-0.8,){/ScalarDash}{1}
\FALabel(10.,14.27)[b]{$\sq$}
\FAVert(6.,10.){0}
\FAVert(14.,10.){1}

\FADiagram{}

\FADiagram{}
\FAProp(0.,10.)(10.,10.)(0.,){/Sine}{0}
\FALabel(5.,8.93)[t]{$V$}
\FAProp(20.,10.)(10.,10.)(0.,){/Sine}{0}
\FALabel(15.,8.93)[t]{$V$}
\FAProp(10.,10.)(10.,10.)(10.,15.5){/ScalarDash}{-1}
\FALabel(10.,16.57)[b]{$\sq$}
\FAVert(10.,10.){1}

\FADiagram{}
\FAProp(0.,10.)(10.,10.)(0.,){/Sine}{0}
\FALabel(5.,8.93)[t]{$V$}
\FAProp(20.,10.)(10.,10.)(0.,){/Sine}{0}
\FALabel(15.,8.93)[t]{$V$}
\FAProp(10.,17.5)(10.,10.)(-0.8,){/ScalarDash}{-1}
\FALabel(14.07,13.75)[l]{$\sq$}
\FAProp(10.,17.5)(10.,10.)(0.8,){/ScalarDash}{1}
\FALabel(5.93,13.75)[r]{$\sq$}
\FAVert(10.,10.){0}
\FAVert(10.,17.5){1}

\FADiagram{}
\end{feynartspicture}
\end{center}
\caption[]{
Generic Feynman diagrams for the vector boson self-energies with
counter\-term insertion.
$V$ denotes either $W$ or $Z$, $\sq$ is either a $\Stop$ or a $\Sbot$,
while $q$ is a $t$ or a $b$}
\label{fig:fdvb1lct}
\end{figure}

In the following sections we describe the necessary ingredients for the
evaluation of these contributions, starting with the relevant sectors
of the MSSM and then turning to the technical evaluation of the
two-loop diagrams and the counterterms.

\subsection{The relevant MSSM sectors}
\label{sec:mssm}

Here we specify the MSSM contributions that are relevant for the 
\order{\alt^2}, \order{\alt \alb}, \order{\alb^2} corrections. 
As explained above, the
calculation involves
the gauge-less limit where $\MW, \MZ \to 0$ 
(keeping $\cw = \MW/\MZ$ fixed). Accordingly, we discuss the
implications of the gauge-less limit for the different sectors of the
MSSM.


\subsubsection{The scalar top and bottom sector}
\label{subsec:squark}

We start with the stop and sbottom sector.
In order to fix the notation, we explicitly list 
their mass matrices in the basis of the interaction eigenstates 
$\StopL, \StopR$ and $\SbotL, \SbotR$:
\BEA
\label{stopmassmatrix}
{\cal M}^2_{\Stop} &=&
  \ML \MstL^2 + \mt^2 + \CZb (\edz - \frac{2}{3} \sw^2) \MZ^2 &
      \mt \Xt \\
      \mt \Xt &
      \MstR^2 + \mt^2 + \frac{2}{3} \CZb \sw^2 \MZ^2 
  \MR, \\
&& \non \\
\label{sbotmassmatrix}
{\cal M}^2_{\Sbot} &=&
  \ML \MsbL^2 + \mb^2 + \CZb (-\edz + \frac{1}{3} \sw^2) \MZ^2 &
      \mb \Xb \\
      \mb \Xb &
      \MsbR^2 + \mb^2 - \frac{1}{3} \CZb \sw^2 \MZ^2 
  \MR 
\EEA
where 
\BE
\mt \Xt \equiv \mt (\At - \mu \CTb) , \quad
\mb\, \Xb \equiv \mb\, (\Ab - \mu \Tb) .
\label{eq:Xtb}
\EE
Here $\At$ denotes the trilinear Higgs--stop coupling, $\Ab$ denotes the
Higgs--sbottom coupling, and $\mu$ is the higgsino mass parameter.

In order to obtain the mass eigenvalues the mass matrix has to be
diagonalized. This is done with the help of the unitary matrix
$U^{\sq}$,
\BE
U^{\sq} {\cal M}^2_{\sq} U^{\sq}{}^\dagger  =  {\rm
  diag}(m_{\sq_1}^2,m_{\sq_2}^2), \quad q = t,b,
\EE
which we will choose real since we consider the MSSM without
$\cp$-violation. 

Furthermore, SU(2) gauge invariance requires the relation
\BE
\MstL^2 = \MsbL^2 
\label{eq:SU2rel}
\EE
between the soft supersymmetry-breaking parameters in the left-handed 
components of the squark doublet. 
Consequently, there are only five
independent parameters in the $\Stop/\Sbot$ sector.  The masses and
mixing angles are connected via the relation 
\BE
\sum_{i=1,2}|U^{\Sbot}_{i1}|^2 \, m_{\Sbot_i}^2=
\sum_{i=1,2}|U^{\Stop}_{i1}|^2 \, m_{\Stop_i}^2\
+\mb^2 - \mt^2 - \MZ^2 \cw^2  \CZb .
\label{SfMassRelation}
\EE

In the gauge-less limit the terms proportional to $\MZ^2$ in the
diagonal entries of the mass matrices and in \refeq{SfMassRelation} vanish.

Except where stated otherwise, we will assume universality
of all three soft supersymmetry-breaking parameters in the diagonal
entries of the stop/sbottom mass matrices,
\BE
\msusy\equiv \MstL = \MstR =\MsbR~.
\EE
The common squark mass scale is denoted as $\msusy$.


\subsubsection{The Higgs sector}
\label{subsec:mssmhiggs}

Contrary to the SM, in the MSSM two Higgs doublets
are required~\cite{hhg}.
At the tree-level, the Higgs sector can be described with the help of two  
independent parameters (besides $g_1$ and $g_2$): the ratio of the two
vacuum expectation values,  
$\tb = v_2/v_1$, and $M_A$, the mass of the $\cp$-odd $A$ boson.
The diagonalization of the bilinear part of the Higgs potential,
i.e.\ the Higgs mass matrices, is performed via orthogonal
transformations with the angle $\al$ for the $\cp$-even part and with
the angle $\be$ for the $\cp$-odd and the charged part.
The mixing angle $\al$ is determined through
\BE
\tan 2\al = \tan 2\be \; \frac{\MA^2 + \MZ^2}{\MA^2 - \MZ^2} ;
\qquad  -\frac{\pi}{2} < \al < 0~.
\label{alphaborn}
\EE
One gets the following Higgs spectrum:
\BEA
\mbox{2 neutral bosons},\, {\cal CP} = +1 &:& h, H \non \\
\mbox{1 neutral boson},\, {\cal CP} = -1  &:& A \non \\
\mbox{2 charged bosons}                   &:& H^+, H^- \non \\
\mbox{3 unphysical Goldstone bosons}      &:& G^0, G^+, G^- .
\EEA
At the tree level, the Higgs boson masses expressed
through $\MZ, \MW$ and $\MA$ are given by
\BEA
\label{mlh}
\Mh^2 &=& \edz \KKL \MA^2 + \MZ^2 - 
          \sqrt{(\MA^2 + \MZ^2)^2 - 4 \MA^2\MZ^2 \CQZb} \KKR, \\
\label{mhh}
\MH^2 &=& \edz \KKL \MA^2 + \MZ^2 + 
          \sqrt{(\MA^2 + \MZ^2)^2 - 4 \MA^2\MZ^2 \CQZb} \KKR, \\
\label{mhp}
\MHp^2 &=& \MA^2 + \MW^2 .
\EEA
We use the Feynman gauge, which implies the mass parameters 
$\MG^2 = \MZ^2$, $\MGp^2 = \MW^2$ for
the unphysical scalars $G^0$ and $G^{\pm}$.

In the gauge-less limit the Higgs sector parameters satisfy the
following relations:
\begin{subequations}
\label{gaugelessHiggs}
\BEA
\MHp^2 &=& \MH^2 = \MA^2~, \label{eq:glmasses}\\
\Sa &=& -\Cb , \quad \Ca = \Sb \label{eq:higgsangles},\\
\MG^2 &=& \MGp^2 = 0 ~. \label{eq:glgoldstones}
\EEA
\end{subequations}
In the gauge-less limit furthermore the mass of the lightest $\cp$-even
Higgs boson vanishes at tree level,
\begin{equation}
\Mh = 0~.
\label{eq:Mh0}
\end{equation}
Because of the accidental cancellation in the SM result for $\MHSM = 0$,
see \refeq{SMMH0}, it is desirable to retain the dependence on $\Mh$ as
much as possible in the MSSM result. In this way the numerical impact of
the $\Mh$-dependence can be studied, which within the MSSM is formally a
higher-order effect. This is particularly interesting in view of the
fact that higher-order corrections to the masses and mixing angles in
the MSSM Higgs sector are sizable, see e.g.\ \citere{habilSH} for
recent reviews. 

We will therefore discuss the implementation of the gauge-less limit in some
detail. In particular, we will investigate in how far a consistent
result for $\De\rho$ can be obtained without imposing \refeq{eq:Mh0}.
We will also briefly discuss the case where \refeq{eq:higgsangles} is
relaxed, see \refse{sec:renorm} below. For 
higher-order corrections in the Higgs sector we use the results
as implemented into the code
{\em FeynHiggs}~\cite{feynhiggs,feynhiggs1.2,mhiggsAEC,feynhiggs2.2}.


\subsubsection{Higgsinos}
\label{subsec:higgsinos}

In the gauge-less limit the contributions from the chargino and neutralino
sector reduce to those of the higgsinos. 

The diagonalization matrices for the chargino and neutralino sectors
in this limit are given by
(see \citere{PomssmRep} for our notation)
\BE
{\bf U} = {\bf V} = \ML 0 & 0 \\ 0 & 1 \MR\, ,\quad
{\bf N} = \frac{1}{\sqrt2}
          \MLv 0 & 0 & 0 & 0 \\
               0 & 0 & 0 & 0 \\
               0 & 0 & 1 &-1  \\
               0 & 0 & 1 & 1 \MR \, ,
\EE
and the corresponding elements of the diagonalized mass matrices are
\BE
m_{\tilde{\chi}^\pm_{i}}=(0,+\mu),\quad
m_{\tilde{\chi}^0_i}=(0,0,+\mu,-\mu).
\EE
All entries corresponding to gauginos are zero since the
gaugino couplings vanish in the gauge-less limit.
Note that the negative sign in $m_{\tilde{\chi}^0_4}$ has to be taken
into account; the physical masses of the charged and neutral
higgsinos are all equal to $+|\mu|$ in the gauge-less limit.

\subsection{Evaluation of the Feynman diagrams}

The amplitudes of all Feynman diagrams shown in 
\reffis{fig:genericsquark},\ref{fig:generichiggsino},\ref{fig:fdvb1lct}
have been created with the 
program {\em FeynArts3}~\cite{feynarts}, making use of the 
MSSM model file~\cite{famssm}.
Dirac algebra and traces have been evaluated using the program
\tc~\cite{2lred},
and the reduction to scalar integrals has been performed in two ways
using either the package {\em TYReduce} \cite{TYReduce} or the routines
built into \tc, based on the reduction method of \citere{2lred}. 
As a result we
obtained the analytical expression for $\De\rho$ depending on the
\onel\ functions $A_0$ and $B_0$~\cite{a0b0c0d0} and on the \twol\
function $T_{134}$~\cite{2lred,t134}. For the further evaluation the
analytical expressions for $A_0$, $B_0$ and $T_{134}$ have been
inserted. 

In addition to the two-loop diagrams, one-loop counterterms
corresponding to the renormalization of divergent one-loop subdiagrams 
have to be taken into account. The whole calculation can be performed
both in dimensional regularization~\cite{dreg} as well as in dimensional
reduction~\cite{dred}. Since no gauge bosons appear in the loops, both
regularization schemes preserve gauge 
invariance and supersymmetry for the present calculation. Therefore
the necessary counterterms correspond to multiplicative
renormalization of the parameters in the MSSM Lagrangian. 

\subsection{Counterterms}

The renormalization constants that are relevant in the gauge-less limit are
\BEA
&\de \mt,\ \de\mb,\
\de\mste^2,\ \de\mstz^2,\
\de\tst,\ \de\msbe^2,\ \de\msbz^2,\ \de\tsb,
&\nonumber\\
&
\de \MA^2,\ \de\tan\beta,\ \de t_{h,H},\
\de\mu,
&
\label{renconstlist}
\EEA
corresponding to the renormalization of the
fermion masses, the parameters of the $\Stop/\Sbot$~sector, the Higgs
sector parameters and tadpoles, and the $\mu$-parameter.
It is not necessary to introduce wave function renormalization constants
for the fermions and scalar fields since they drop out in the sum over
all diagrams.

There are two possible ways to obtain the counterterm contributions:
\begin{enumerate}
\item Generate and evaluate one-loop diagrams with insertions of
  counterterm vertices, as depicted generically in
  \reffi{fig:fdvb1lct}. In order to generate these diagrams, the
  required counterterm Feynman rules had to be added to the {\em
  FeynArts} MSSM model file. In the explicit evaluation of the
  counterterm diagrams it turned out that the renormalization
  constants $ \de \MA^2,\ \de\tan\beta,\ \de t_{h,H},\ \de\mu$,
  corresponding to the Higgs/higgsino sector, drop out. Only the
  quark and squark mass and mixing renormalization constants
  contribute.
\item The renormalization transformation
  $\la_i\to\la_i+\de\la_i$ for each parameter $\la_i$
  appearing in eq.\ (\ref{renconstlist}) is
  performed directly in the one-loop result 
  $\De\rho_{\rm 1-loop}^{\rm SM+SUSY}(\la_i)$.
  The counterterm contribution for the two-loop calculation is then
  obtained by expanding $\De\rho_{\rm 1-loop}^{\rm SM+SUSY}(\la_i+\de\la_i)$
  to first order in the $\de\la_i$, where the contributions to 
  $\De\rho_{\rm 1-loop}^{\rm SM+SUSY}$ have been given in 
  \refeqs{eq:delrhoSM1l}, (\ref{delrhoMSSM1l}). 
  In this setup it is obvious that
  the renormalization constants 
  $ \de \MA^2,\ \de\tan\beta,\ \de t_{h,H},\ \de\mu$ do not contribute,
  since the one-loop result in the gauge-less limit consists only of the
  quark and squark loop contributions and therefore does not depend on
  the Higgs-sector parameters.
  Accordingly, the counterterm contributions for the
  two-loop calculation, $\De\rho_{\rm ct}$, can be written as
\BE
\De\rho_{\rm ct}=
\sum_{{f}={t},{b}}
\left(\de m_f\partial_{m_f}+\sum_{i=1,2}\de m^2_{\tilde{f}_i}\partial_{m^2_{\tilde{f}_i}}
+\sum_{i,j=1,2}\de U^{\tilde{f}}_{ij}\partial_{U^{\tilde{f}}_{ij}}
\right) \De\rho^{\rm SM+SUSY}_{\rm 1-loop}.
\EE
\end{enumerate}
In order to have a non-trivial check of the counterterm contributions, we
implemented them using both approaches and found agreement in the
final result.

\smallskip
Due to supersymmetry and SU(2) gauge invariance, see \refeq{eq:SU2rel}, 
there are only five independent parameters in the $\Stop/\Sbot$
sector, leading to \refeq{SfMassRelation}.
As a consequence, not all the parameters appearing in
\refeq{SfMassRelation} can be renormalized independently. Choosing
$\msbe^2$ as the dependent parameter, its counterterm $\de \msbe^2$ can
be expressed in terms of the other counterterms. In the gauge-less
limit the relation reads
\begin{align}
\non
\de \msbe^2 &= 
\frac{1}{|U^{\Sbot}_{11}|^2} \Bigl(
\sum_{i=1,2}|U^{\Stop}_{i1}|^2 \de m_{\Stop_i}^2
%
 - |U^{\Sbot}_{21}|^2 \de \msbz^2
 + 2\, U^{\Stop}_{11}U^{\Stop}_{21} 
       (\mste^2 -\mstz^2)\de u^{\Stop}_{12} \\[1.5mm]
& \quad\  \label{deltamsb1}
 -  2\, U^{\Sbot}_{11}U^{\Sbot}_{21} (\msbe^2 -\msbz^2)\de u^{\Sbot}_{12}
-  2 \mt\, \de\mt + 2 \mb\, \de \mb   \Bigr)~,
\end{align}
where the renormalization transformation of the mixing matrix is
defined as 
\BE
U^{\sq}_{ij} \to U^{\sq}_{ij} +\de U^{\sq}_{ij} \, ,\quad
\de U^{\sq}_{ij} = \de u^{\sq}_{ik}U^{\sq}_{kj} \, ,\quad
\de u^{\sq}_{ij}=-\de u^{\sq}_{ji} .
\EE

\smallskip
In order to define the renormalization constants one has to choose a
renormalization scheme. For the SM fermion masses $m_{t,b}$ we always
choose the on-shell scheme. This yields for the top mass counterterm
\begin{align}\label{dmt}
\de\mt = \edz \mt \bigl [\re \Si_{{t}_L} (\mt^2) +
                         \re \Si_{{t}_R} (\mt^2) +
                        2\re \Si_{{t}_S} (\mt^2)
\bigr]\; ,
\end{align}
with the scalar coefficients of the
unrenormalized top-quark self-energy, $\Si_t (p)$,
in the Lorentz decomposition
\begin{align}
\label{Fermionselbstenergiezerlegung}
 \Si_t (p) &= \pslash \OM
\Si_{{t}_L} (p^2) + \pslash \OP
\Si_{{t}_R} (p^2) + \mt \Si_{{t}_S} (p^2)\; ,
\end{align}
and analogously for the bottom mass counterterm
(in order to take higher-order QCD
corrections into account, we use an effective bottom quark mass value of 
$\mb = 3 \gev$). 
For the five independent $\Stop/\Sbot$
sector parameters we choose either the on-shell
\cite{MSSMOS,mhiggsFDalbals} or the \drbar\  scheme. The precise
definitions will be given in the following section.


\section{Renormalization prescriptions and result for
\boldmath{$\De\rho$}}
\label{sec:renorm}

As explained above, the strict implementation of the gauge-less limit 
in the evaluation of the \order{\alt^2}, \order{\alt \alb},
\order{\alb^2} contributions to $\De\rho$ would imply that the mass of
the lightest $\cp$-even Higgs boson has to be set to zero, see
\refeq{eq:Mh0}. In the SM case, where $\MHSM$ is a free parameter,
it turned out that the two-loop Yukawa contribution to $\De\rho$ yields
a much better approximation of the full electroweak two-loop corrections
to the EWPO for (realistic) non-zero values of $\MHSM$ than in the limit
$\MHSM = 0$. It is therefore of interest to investigate the impact of 
non-zero values of $\Mh$ also for the MSSM, where $\Mh$ is a dependent
quantity that is determined by the other supersymmetric parameters.

It has been observed already in \citere{drMSSMgf2A} that
the pure fermion contributions of class $(q)$ (see
\reffi{fig:samplediagrams}) may consistently be obtained 
even if \refeq{eq:Mh0} is not employed. 
In this section we discuss this issue in detail and explain the physical
origin of this behavior. Based on this result we show how the
calculation of all three classes of contributions to $\De\rho$,
i.e.\ $(q)$, $(\sq)$, and $(\tilde{H})$, can be organized in such
a way that $\Mh$ can be set to its true MSSM value essentially
everywhere. We will use the resulting expression in order to study the 
numerical effect of non-zero $\Mh$ values for the new corrections
calculated in this paper, namely the squark contribution $(\sq)$
and the higgsino contribution $(\tilde{H})$.


\subsection{Higgs sector}
\label{HiggsSecRen}

In order to discuss the implementation of the gauge-less limit
it is useful to compare the MSSM case
with the one of a general two-Higgs-doublet model (2HDM).
In the following ``2HDM'' is to be understood as a two-Higgs-doublet
model including squarks and higgsinos, but without any supersymmetric 
relations imposed on them.
The MSSM can be regarded as a special case of a 2HDM, with
supersymmetric relations for
couplings and masses. In the 2HDM without these coupling
relations, $\De\rho$ is also well-defined and can be
calculated at \order{\alt^2}, \order{\alt\alb},
\order{\alb^2} in the gauge-less limit. The corresponding two-loop
diagrams are identical to the diagrams of the classes $(q)$,
$(\sq)$, $(\tilde{H})$ in the MSSM. However, in contrast to the MSSM, the
Higgs-boson masses in the 2HDM are independent parameters 
and do not have to obey  
\refeqs{gaugelessHiggs}, (\ref{eq:Mh0}) in the gauge-less limit. 

The essential difference between the MSSM and the 2HDM case concerns
the renormalization and counterterm contributions.
Restricting ourselves in a first step to class $(q)$, these contributions can
be decomposed in the MSSM and the 2HDM as
\BEA
\label{dominik1}
\De\rho^{(q)}_{\rm MSSM} &=& \De\rho^{(q)}_{\rm 2-loop} +
\De\rho^{(q)}_{tb-{\rm ct}} \\
\label{dominik2}
\De\rho^{(q)}_{\rm 2HDM} &=& \De\rho^{(q)}_{\rm 2-loop} +
\De\rho^{(q)}_{tb-{\rm ct}} + \De\rho^{(q)}_{H-{\rm ct}}~,
\EEA
respectively. Here the two-loop
contribution $\De\rho^{(q)}_{\rm 2-loop}$ and the
counterterm contributions from the $t/b$~doublet
$\De\rho^{(q)}_{tb-{\rm ct}}$ are identical in the two models, 
while the Higgs sector counterterm contribution 
$\De\rho^{(q)}_{H-{\rm ct}}$ appears only in the 2HDM result.

As mentioned above, in the MSSM there are no one-loop contributions 
from the
Higgs sector and correspondingly no Higgs sector counterterm
contributions at \order{\alt^2}, \order{\alt\alb},
\order{\alb^2}. In the 2HDM, the Higgs sector one-loop contribution to
$\De\rho$ reads
\begin{align}
\De\rho^{\rm 2HDM}_{\rm 1-loop} = \frac{g^2}{128\pi^2 \MW^2}
\Big[&
F_0(M_{H^\pm}^2,\MA^2)
\nonumber\\&
+\sin^2({\beta-\alpha})\left(
F_0(M_{H^\pm}^2,M_{H}^2)-F_0(\MA^2,M_{H}^2)\right)
\nonumber\\&
+ \cos^2({\beta-\alpha})\left(
F_0(M_{H^\pm}^2,M_{h}^2)-F_0(\MA^2,M_{h}^2)\right)\Big]
\label{THDMHiggs1L}
\end{align}
in the gauge-less limit. 
Note that this contribution indeed vanishes if
the MSSM gauge-less 
limit relations (\ref{gaugelessHiggs}) hold. 
The counterterm contribution from the Higgs sector 
at the two-loop level can be obtained from this
expression as
\begin{align}
\De\rho^{(q)}_{H-{\rm ct}} \; \equiv \; 
\left(\sum_{\phi=h,H,A^0,H^\pm}\de M_\phi^2 \, \partial_{M_\phi^2}
+\de \tan\beta \, \partial_{\tan\beta} 
+\de \sin\alpha \, \partial_{\sin\alpha}\right)
\De\rho^{\rm 2HDM}_{\rm 1-loop}.
\label{THDMHiggsCT}
\end{align}

Since $F_0(x,y)$ and $\partial_xF_0(x,y)$ vanish in the limit $x=y$,
we find that $\De\rho^{(q)}_{H-{\rm ct}}$
vanishes if 
\BE
M_H=M_{H^\pm}= \MA, \quad \cos(\beta-\alpha)=0, \quad M_h=\mbox{arbitrary}
\label{NecessaryRelations}
\EE
or
\BE
M_{H^\pm}= \MA, \quad \de M_{H^\pm}^2=\de\MA^2, \quad
M_h, \MH, \al =\mbox{arbitrary}.
\label{NecessaryRelations2}
\EE

The relations in \refeq{NecessaryRelations} are the same as the
constraints imposed by the gauge-less limit in the
MSSM except for the fact that $M_h=0$ is not necessary.
The observation made in \citere{drMSSMgf2A} that the class $(q)$ contributions
to $\De\rho$ can be evaluated in the MSSM in a meaningful way for 
non-zero values of $\Mh$ can be understood from
\refeq{NecessaryRelations}. 
For class $(q)$ the two-loop diagrams
and the fermion sector counterterms are identical in the MSSM and the
2HDM. If the relations in
\refeq{NecessaryRelations} hold, 
$\De\rho^{(q)}_{H-{\rm ct}} = 0$ in \refeq{dominik2}, so that
\refeq{dominik1} and \refeq{dominik2} become identical. 
Thus, the calculations in the MSSM and the 2HDM are
the same in this case. This means that the result of class $(q)$
derived in the MSSM 
for non-zero $\Mh$ is well-defined and consistent, as it corresponds to 
a certain special case of the general 2HDM result.


\subsection{Inclusion of the $\Stop/\Sbot$ sector in the on-shell scheme}

For the full set of contributions to $\De\rho$, also the sfermion
diagrams of class $(\tilde q)$ and the higgsino diagrams of class
$(\tilde H)$ have to be taken into account. In the following, as
explained above, we consider 
a 2HDM including also stops, sbottoms and higgsinos (although without any
supersymmetric relations). The analogy of the calculation in the
MSSM and the 2HDM does no longer hold, since the
sfermion sector renormalization differs in the two models. 

As discussed above, supersymmetry and SU(2) gauge invariance imply that
not all parameters in the squark sector can be renormalized
independently in the MSSM. 
Choosing $\msbe$ as the dependent mass in the MSSM, 
its renormalization constant
$\de \msbe^2$ is given by \refeq{deltamsb1}, and no
independent renormalization condition can be imposed on it. We will
refer to the expression for $\de \msbe^2$ in terms of the
other counterterms of the fermion and sfermion sector as given in
\refeq{deltamsb1} as the ``symmetric'' renormalization, 
$\de \msbe^2\big|_{\rm symm}$.

In the on-shell scheme \cite{MSSMOS}, the three other squark masses 
$m_{\tilde{t}_{1,2},\tilde{b}_2}$
are defined as pole masses, and the mixing angle counterterms can be 
defined via on-shell mixing self-energies:
\begin{align}
\de \msfi^2 & =  {\rm Re} \, \Sigma_{\tilde{f}_i}(\msfi^2)
&&\mbox{ for }\tilde{f}_i=\tilde{t}_{1,2},\tilde{b}_2;\\
\de u^{\tilde{f}}_{12} & = 
\frac{{\rm Re} \, \Sigma_{\tilde{f}_1\tilde{f}_2}(m_{\tilde{f}_1}^2)
+{\rm Re} \, \Sigma_{\tilde{f}_1\tilde{f}_2}(m_{\tilde{f}_2}^2)}
{2(m_{\tilde{f}_1}^2-m_{\tilde{f}_2}^2)}
&&\mbox{ for }\tilde{f}=\tilde{t},\tilde{b}.
\end{align}

In a 2HDM with squarks, on the other
hand, an on-shell renormalization can be applied for all four squark
masses. In this case $\de \msbe^2$ is given by 
\BE
\de \msbe^2\big|_{\rm OS} =
{\rm Re} \,
\Sigma_{\tilde{b}_1}(\msbe^2) ,
\label{msb1OS}
\EE
where $\Sigma_{\tilde{b}_1}$ is the 
$\tilde{b}_1$ self-energy.
Hence there is a mass shift
\BE
\De \msbe^2=\de
\msbe^2\big|_{\rm symm} - \de \msbe^2\big|_{\rm OS}
\label{massshift}
\EE
at the one-loop level between the
mass parameter $\msbe^2$ as given by the ``symmetric''
renormalization and the physical pole mass.

For the class $(\sq, \tilde H)$ the decomposition of $\De\rho$ in the
two models is given by
\BEA
\label{dominik3}
\De\rho^{(\sq,\tilde H)}_{\rm MSSM} &=& 
\De\rho^{(\sq, \tilde H)}_{\rm 2-loop} +
\De\rho^{(\sq, \tilde H)}_{tb-{\rm ct}} +
\De\rho^{(\sq, \tilde H)}_{\Stop\Sbot-{\rm ct, \; symm}} \\
\label{dominik4}
\De\rho^{(\sq, \tilde H)}_{{\rm 2HDM}} &=& 
\De\rho^{(\sq, \tilde H)}_{\rm 2-loop} + 
\De\rho^{(\sq, \tilde H)}_{tb-{\rm ct}} +
\De\rho^{(\sq, \tilde H)}_{\Stop\Sbot-{\rm ct, \; full \; OS}} + 
\De\rho^{(\sq, \tilde H)}_{H-{\rm ct}}~.
\EEA
Here $\De\rho^{(\sq, \tilde H)}_{\Stop\Sbot-{\rm ct, \; symm}}$
corresponds to the ``symmetric'' renormalization of the squark sector in
the MSSM described above. 
$\De\rho^{(\sq, \tilde H)}_{\Stop\Sbot-{\rm ct, \; full \; OS}}$
denotes the contribution from the full on-shell renormalization of all
squarks. 
As one can see from \refeqs{dominik3} and (\ref{dominik4}) the MSSM result
differs from the 2HDM result even for the case where
$\De\rho^{(q)}_{H-{\rm ct}} = 0$. The MSSM result therefore does not
correspond to a special case of the 2HDM expression.


\subsection{Result for $\De\rho$ in the on-shell scheme}
\label{subsec:delrhoOS}

The total result for $\De\rho$ at \order{\alt^2, \alt\alb, \alb^2} in
the MSSM is given by the sum of \refeqs{dominik1} and (\ref{dominik3}),
\BE
\De\rho^{(q,\sq,\tilde H)} = \De\rho^{(q)}_{\rm MSSM} 
+ \De\rho^{(\sq, \tilde H)}_{\rm MSSM}~. 
\EE
As discussed above, in $ \De\rho^{(\sq, \tilde H)}_{\rm MSSM}$
the ``symmetric'' renormalization
in the sfermion sector has to be applied, leading to the relations
(\ref{SfMassRelation}), (\ref{deltamsb1})
for $\msbe^2$ and $\de \msbe^2$.
The contribution of class $(\sq, \tilde H)$ can be rewritten by using
the mass shift as defined in \refeq{massshift}
(see also \citere{dr2lA}), leading to the expression
\BE
\De\rho^{(q, \tilde q, \tilde H)} =
\De\rho^{(q)}_{\rm MSSM} + 
\De\rho^{(\tilde q, \tilde H)}_{\rm MSSM, \; full \; OS} + 
\De \msbe^2\partial_{\msbe^2}\De\rho_{\rm 1-loop}^{\rm SUSY} ~,
\label{eq:finalres}
\EE
where $\De\rho^{(\tilde q, \tilde H)}_{\rm MSSM, \; full \; OS}$ is
given by
\BE
\De\rho^{(\tilde q, \tilde H)}_{\rm MSSM, \; full \; OS} =
\De\rho^{(\sq, \tilde H)}_{\rm 2-loop} +
\De\rho^{(\sq, \tilde H)}_{tb-{\rm ct}} +
\De\rho^{(\sq, \tilde H)}_{\Stop\Sbot-{\rm ct, \; full \; OS}}~.
\label{delrhofullOS}
\EE
$\De\rho^{(q)}_{\rm MSSM}$ is the result for the fermion-loop contributions as
obtained in \citere{drMSSMgf2A} (employing an on-shell renormalization
of the fermion masses and inserting non-zero values for $\Mh$).
$\De\rho^{(\tilde q, \tilde H)}_{\rm MSSM, \; full \; OS}$ is the
contribution of the squark and higgsino diagrams
obtained by renormalizing {\em all\/} sfermion masses, i.e.\ 
including $\msbe^2$, on-shell, 
while the last term in \refeq{eq:finalres}
is a symmetry-restoring contribution involving $\De \msbe^2$.
The one-loop sfermion contribution $\De\rho_{\rm 1-loop}^{\rm SUSY}$
has been defined in \refeq{delrhoMSSM1l}. 

Comparing \refeqs{delrhofullOS} and (\ref{dominik4}) shows that the 
``full OS'' contribution in \refeq{delrhofullOS} is UV-finite already
for the partial gauge-less limit of \refeq{NecessaryRelations},
according to the discussion of the previous two subsections.

Consistency requires that the mass shift in \refeq{massshift} has to be
UV-finite as well. One can easily check that this requires to take into account
both squark/Higgs and quark/higgsino loops. Correspondingly,
because of the necessity of this shift only the sum of the
$(\sq)$ and $(\tilde{H})$ contributions to $\De\rho$ is
physically meaningful in the MSSM. 

Moreover, the mass shift $\De \msbe^2$ is only finite in the
gauge-less limit, i.e.\ it can only consistently be evaluated if all the
gauge-less limit relations (\ref{eq:glmasses})--(\ref{eq:glgoldstones})
{\em and \/} $\Mh = 0$, \refeq{eq:Mh0}, are used. The last term in 
\refeq{eq:finalres} can therefore only be obtained in the approximation
where $\Mh = 0$. 

The expression in \refeq{eq:finalres} represents the main result of this
paper. For the first term on the right-hand side of \refeq{eq:finalres},
$\De\rho^{(q)}_{\rm MSSM}$, we keep the full dependence on $\Mh$. As
explained above, this is possible because this term is not affected by
the renormalization in the sfermion sector. For the second term, 
$\De\rho^{(\tilde q, \tilde H)}_{\rm MSSM, \; full \; OS}$, we will keep the 
$\Mh$-dependence as well and compare to the strict gauge-less limit case 
where $\Mh = 0$. The mass shift $\De \msbe^2$ entering the
last term in \refeq{eq:finalres} is evaluated for $\Mh = 0$.

The last term in \refeq{eq:finalres} can be expressed using \refeqs{dominik3}
and (\ref{dominik4}) as
\BE
\De\msbe^2 \dd_{\msbe^2} \De\rho^{\rm SUSY}_{\rm 1-loop} =
\De\rho^{(\sq, \tilde H)}_{\Stop\Sbot-{\rm ct, \; symm}}\big|_{\Mh = 0} -
\De\rho^{(\sq, \tilde H)}_{\Stop\Sbot-{\rm ct, \; full \; OS}}\big|_{\Mh = 0}~.
\label{massshift2}
\EE
Correspondingly the full result can be rewritten as
\BEA
\De\rho^{(q,\sq, \tilde H)} &=& 
\De\rho^{(q)}_{\rm MSSM} + 
\De\rho^{(\sq, \tilde H)}_{\rm 2-loop} +
\De\rho^{(\sq, \tilde H)}_{tb-{\rm ct}} +
\De\rho^{(\sq, \tilde H)}_{\Stop\Sbot-{\rm ct, \; full \; OS}} \non \\
\mbox{}&& +
\KKL
\De\rho^{(\sq, \tilde H)}_{\Stop\Sbot-{\rm ct, \; symm}}\big|_{\Mh =
  0} -
\De\rho^{(\sq, \tilde H)}_{\Stop\Sbot-{\rm ct, \; full \; OS}}\big|_{\Mh = 0}
\KKR~.
\label{finalres2}
\EEA
All contributions in the first line of \refeq{finalres2} can be
evaluated by keeping the full $\Mh$ dependence. 
For the other parameters of the Higgs sector we impose the gauge-less
limit as specified in (\ref{eq:glmasses})--(\ref{eq:glgoldstones}).

As a result of \refeq{NecessaryRelations2}, the gauge-less limit
can be relaxed in another way. If the sum of all contributions
$(q,\sq,\tilde{H})$ is considered, the relation $\de \MHp^2=\de
\MA^2$ in \refeq{NecessaryRelations2} is valid. As a consequence,
in the evaluation of the first line of \refeq{finalres2} it is not even
necessary to use the gauge-less limit for $\sin\al$ and $\MH$.
Instead, $\sin\al$ and $\MH$ can be set to their true
MSSM values.
We will discuss the case where the gauge-less limit is relaxed also
for these two parameters below.


\subsection{Renormalization in the \drbar\ scheme }

As an alternative to the on-shell scheme in the squark sector, we also
consider the \drbar\ scheme. In this scheme
the counterterms of the soft
supersymmetry-breaking parameters are defined to be pure
divergences. The squark mass and mixing angle counterterms receive
finite contributions corresponding to $m_{t,b}$ in the squark mass
matrices (\ref{stopmassmatrix}), (\ref{sbotmassmatrix}):
\BEA
\label{dominikA}
\de \msfi^2\big|_{\rm fin} & = & 
\left(U^{\tilde{f}}\de{\cal
  M}^2_{\tilde{f}}U^{\tilde{f}}{}^\dagger\right)_{ii}
\mbox{ for }\tilde{f}_i=\tilde{t}_{1,2},\tilde{b}_{1,2};\\
\label{dominikB}
\de u^{\tilde{f}}_{12}\big|_{\rm fin} & = & 
\frac{\left(U^{\tilde{f}}\de{\cal
  M}^2_{\tilde{f}}U^{\tilde{f}}{}^\dagger\right)_{12}}
{m_{\tilde{f}_1}^2-m_{\tilde{f}_2}^2},\\
\label{dominikC}
\de{\cal   M}^2_{\tilde{f}}\big|_{\rm fin}  & = & 
\de m_{f}\big|_{\rm fin}
\left(\begin{array}{cc}2m_f & X_f\\X_f & 2m_f\end{array}\right).
\EEA
The result for $\De\rho^{(q,\sq,\tilde H)}$ in the \drbar\ scheme
follows from \refeq{finalres2} by replacing 
$\De\rho^{(\sq, \tilde H)}_{\Stop\Sbot-{\rm ct, \; full \; OS}}$ by
the corresponding counterterm resulting from
\refeqs{dominikA}--(\ref{dominikC}). As a consequence, the terms in the
second line of \refeq{finalres2} vanish. 
The results in the \drbar\ scheme depend on the renormalization scale 
$\mu^{\drbarm}$.


\section{Numerical analysis}
\label{sec:numanal}

In this section the numerical effect of the electroweak two-loop 
correction \refeq{eq:finalres}, or equivalently \refeq{finalres2},
is analyzed, using the formulas in 
\refeq{precobs1} to obtain the corresponding shift in $\MW$ and $\sweff$. 
In addition to the full MSSM 
correction resulting from $\De\rho^{(q,\sq,\tilde{H})}$, we also present
the effective change compared to the SM result (where the SM Higgs boson mass
has been set to $\Mh$).
This effective change can be decomposed into the contribution from class
$(q)$ and from classes $(\sq,\tilde H)$.
The contribution from class $(q)$, which was studied in
\citere{drMSSMgf2A}, reads
\BE
\De\rho^{(q)}({\rm MSSM-SM}) = \De\rho^{(q)}_{\rm MSSM} -
\De\rho^{\SM,\al_t^2}_{\rm 2-loop}(\MHSM = \Mh), 
\label{eq:delrhodiff}
\EE
where $\De\rho^{\SM,\al_t^2}_{\rm 2-loop}$ has been given in
\refeq{drSMgf2}.
The contribution from classes $(\sq,\tilde H)$ is given by
\BE
\De\rho^{(\sq,\tilde H)} = 
\De\rho^{(\tilde q, \tilde H)}_{\rm MSSM, \; full \; OS} +
\De \msbe^2\partial_{\msbe^2}\De\rho_{\rm 1-loop}^{\rm SUSY}~,
\label{massshiftincl}
\EE
where $\Mh=0$ is used in the second term. Here and in the following we
drop the subscript ``MSSM'' for simplicity. 

As SM input parameters we use the values
$\mt= 178.0 \gev, 
\mb = 3 \gev$.
The bottom quark mass is to be understood as an effective bottom quark
mass, taking into account higher-order QCD corrections.


\subsection{Impact of relaxing the gauge-less limit for $\Mh$ and $\sin\alpha$}
\label{sec:51}

In the first step we study the impact of
evaluating $\De\rho^{(\tilde q, \tilde H)}_{\rm MSSM, \; full \; OS}$ 
(see \refeq{massshiftincl}) for the true value of the lightest MSSM
Higgs-boson mass $\Mh$ rather than for $\Mh = 0$. Accordingly, we
compare the effect on the EWPO resulting from 
$\De\rho^{(q)}(\Mh)+\De\rho^{(\sq,\tilde{H})}(\Mh)$ and
$\De\rho^{(q)}(\Mh)+\De\rho^{(\sq,\tilde{H})}(0)$. 

\begin{figure}[htb!]
\begin{center}
\epsfig{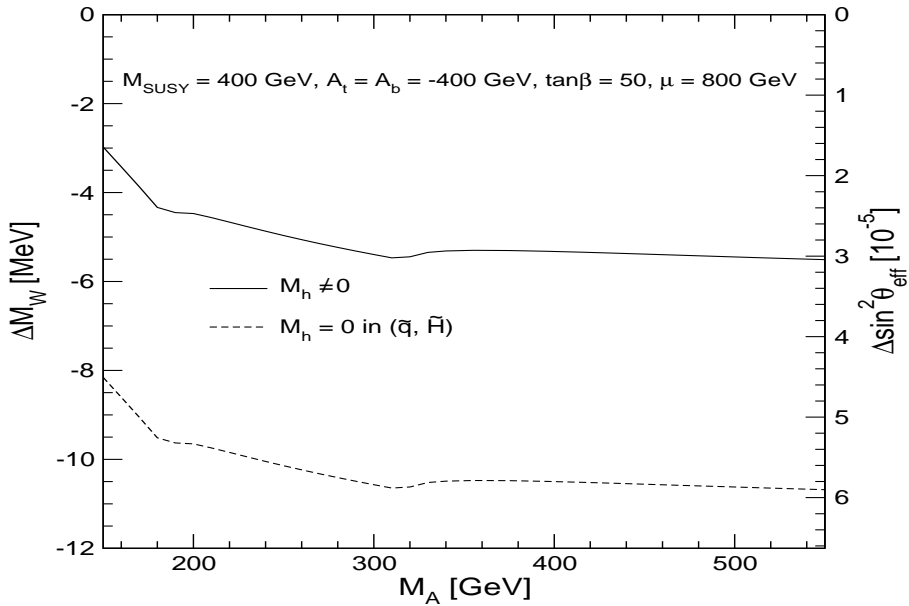}
\caption{%
$\De\MW$ and $\De\sweff$ are shown as a function for $\MA$ for 
the case where the full dependence on the mass of the light $\cp$-even
Higgs boson is kept,
$\De\rho^{(q)}(\Mh)+\De\rho^{(\sq,\tilde{H})}(\Mh)$, and for the case
where the strict gauge-less limit for $\Mh$ has been applied in the squark 
and higgsino contributions,
$\De\rho^{(q)}(\Mh)+\De\rho^{(\sq,\tilde{H})}(0)$.
}
\label{fig:Mheq0}
\end{center}
\end{figure}

We have investigated
the numerical effect of keeping the dependence on $\Mh$ in the squark
and higgsino contributions for various MSSM scenarios. \reffi{fig:Mheq0}
shows an example where the numerical impact on the prediction of $\MW$
and $\sweff$ is quite sizable.
The EWPO are given as a function of $\MA$ with 
$\msusy = -A_{t,b} = 400 \gev$, $\mu = 800 \gev$ and $\tb = 50$.
The effect of keeping a non-vanishing value of $\Mh$ in the squark and
higgsino contributions amounts to about $+5 \mev$ in $\MW$ and 
$-3 \times 10^{-5}$ to $\sweff$ for all considered $\MA$ values.
The effects for other MSSM scenarios are typically smaller than for the
example shown in \reffi{fig:Mheq0}. Unless otherwise stated, we will
always keep the full $\Mh$ dependence in the results shown below. The
difference between the result with and without the $\Mh$ dependence
can be employed for
estimating the residual theoretical uncertainties from unknown
higher-order corrections, see the discussion in \refse{subsec:intrinsicfuture}
below.

\begin{figure}[t!]
\begin{center}
\epsfig{figure=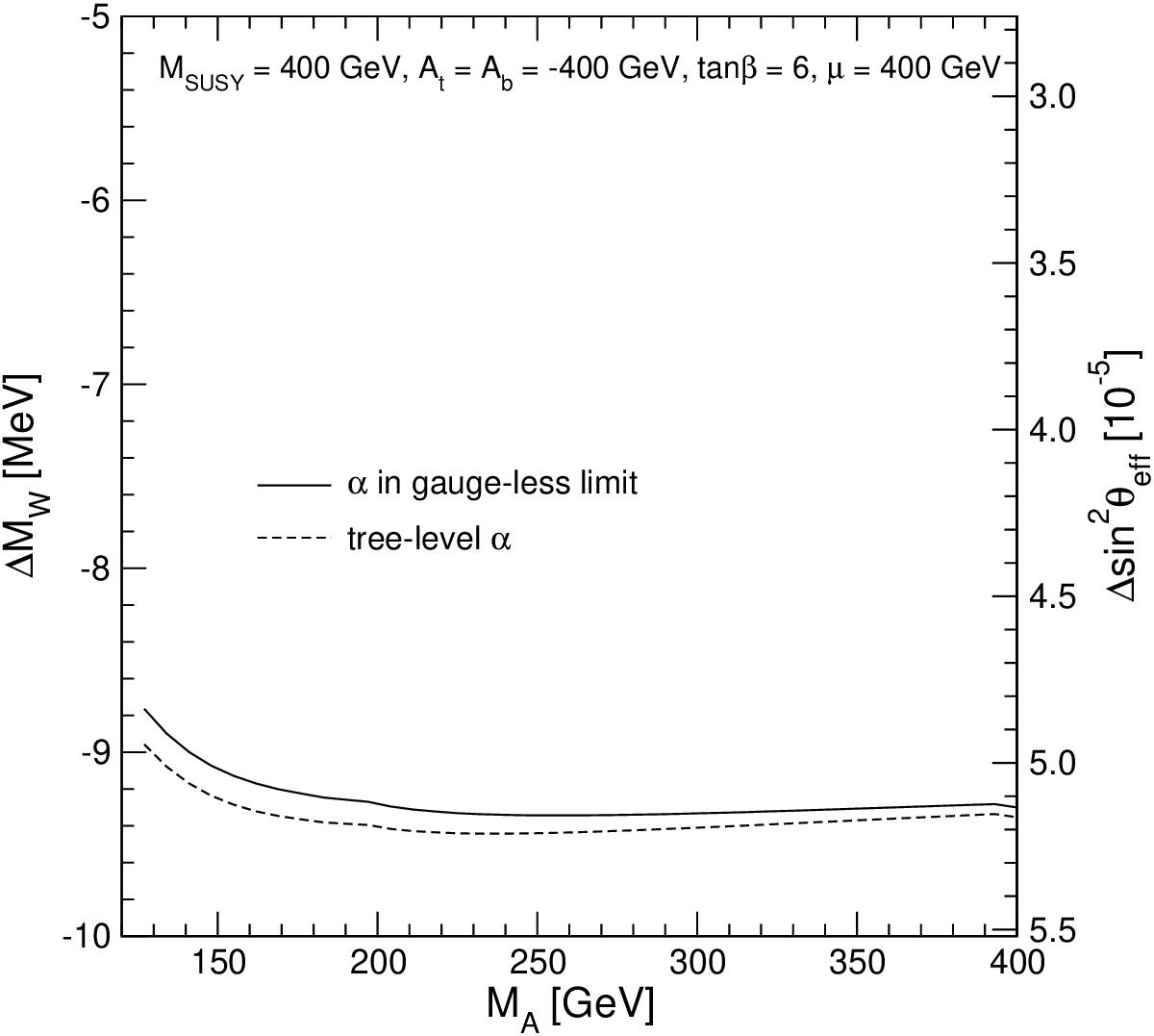,width=12cm,height=8cm}
\caption{%
$\De\MW$ and $\De\sweff$ are shown for the case where the Higgs mixing
  angle $\al$ obeys either the full tree-level relation,
  \refeq{alphaborn}, or is fixed by the
  gauge-less limit,
  \refeq{eq:higgsangles}.
}
\label{fig:alpha}
\end{center}
\end{figure}

\reffi{fig:alpha} illustrates the numerical effect of relaxing the
gauge-less limit on $\sin\al$. As discussed at the end of
\refse{subsec:delrhoOS}, the
sum of the contributions of classes $(q,\sq,\tilde H)$ can be evaluated
in a meaningful way even
if $\sin\al$ and $\MH$ are set to their true values in the MSSM
instead of their values in the gauge-less limit.
Since the corresponding shift in $\MH$ is
usually quite small~\cite{feynhiggs} we do not analyze the effects
arising from different choices for $\MH$ and use its gauge-less value
throughout the paper. The situation is different for the
Higgs mixing angle~$\al$. Here the full tree-level 
value $\sin\al^{\rm full}$ as given in \refeq{alphaborn} 
can significantly deviate from its gauge-less value, 
$\sin\al^{\rm gl}=-\cos\be$. 
\reffi{fig:alpha} shows the results for $\De\MW$ and $\De\sweff$ based
on $\sin\al^{\rm full}$ and $\sin\al^{\rm gl}$.
The parameters are chosen in such a way as to maximize the influence of
$\sin\al^{\rm full}$ vs.\ $\sin\al^{\rm gl}$. The
value $\tb = 6$ is rather small, and e.g.\ together with 
$\MA=150\gev$ it leads to $\sin\al^{\rm full}=-0.31$ and 
$\sin\al^{\rm gl}=-0.16$. For $\msusy = \mu = 400 \gev$
and $A_{t,b}=-800$ this parameter set is in agreement
with all experimental constraints from Higgs boson
searches~\cite{LEPHiggsSM,LEPHiggsMSSM} and 
$b$-physics~\cite{pdg}. \reffi{fig:alpha} shows that even in this
scenario the numerical effect of relaxing the gauge-less limit on 
$\sin\al$ is negligible. We have
checked that this holds in general. 
In particular for larger $\tb$ and/or $\MA$ the effect is
even smaller. Therefore we will always set $\sin\al$ to
$\sin\al^{\rm gl}$ in the following.

\subsection{Dependence on supersymmetric parameters}
\label{subsec:ewponum}

\begin{figure}[th!]
\begin{center}
\epsfig{figure=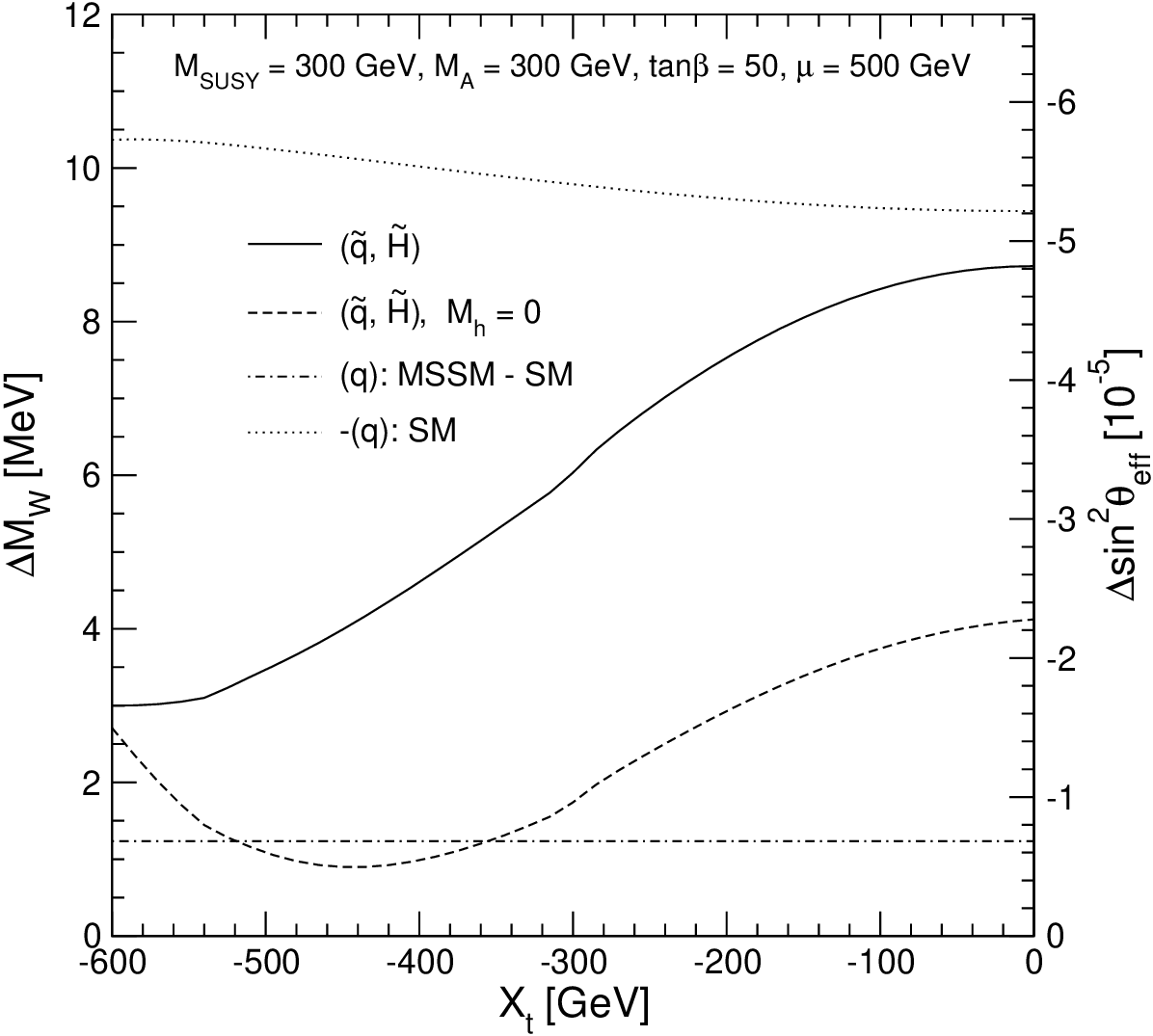,width=12cm,height=8.2cm}
\caption{%
$\De\MW$ and $\De\sweff$ are shown as a function of $\Xt$ in a
  scenario with large $\tan\beta$.
The two-loop contribution involving squarks and higgsinos,
$\De\rho^{(\sq,\tilde{H})}$, is shown for the correct MSSM value of
$\Mh$ and for $\Mh = 0$.
For the class $(q)$ the effective change from the
SM to the MSSM is shown and compared with the pure SM contribution
(with the sign reversed for better visibility). 
}
\label{fig:LargeTB}
\end{center}
\end{figure}

\begin{figure}[th!]
\begin{center}
\epsfig{figure=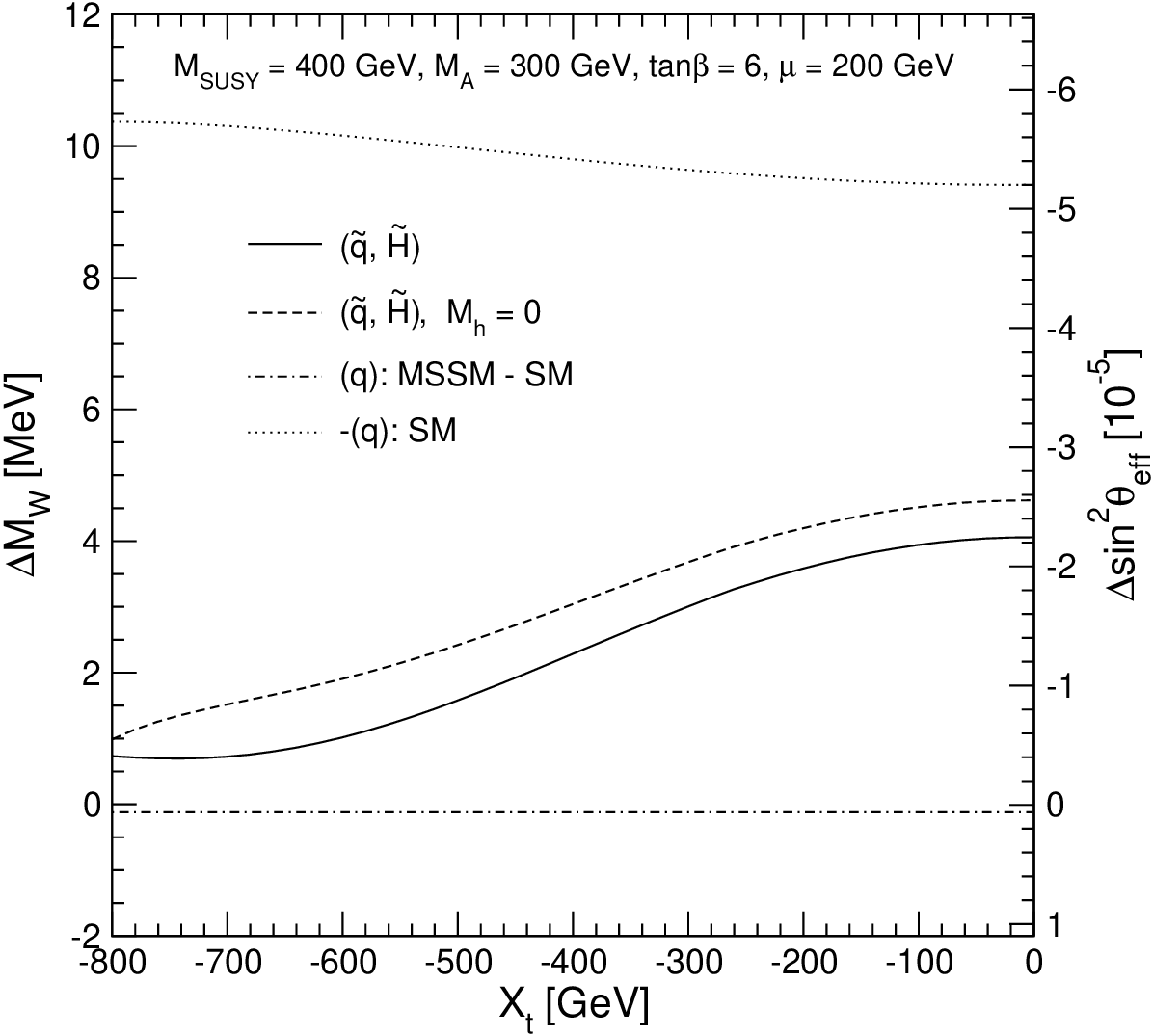,width=12cm,height=8.5cm}
\caption{%
$\De\MW$ and $\De\sweff$ are shown as a function of $\Xt$ in a
  scenario with small $\mu$ and $\tan\beta$.
The two-loop contribution involving squarks and higgsinos,
$\De\rho^{(\sq,\tilde{H})}$, is shown for the correct MSSM value of
$\Mh$ and for $\Mh = 0$.
For the class $(q)$ the effective change from the
SM to the MSSM is shown and compared with the pure SM contribution
(with the sign reversed for better visibility). 
}
\label{fig:SmallMUE}
\end{center}
\end{figure}

In \reffis{fig:LargeTB}, \ref{fig:SmallMUE} and \ref{fig:Standard} we
explore the numerical impact of
$\De\rho^{(q,\sq,\tilde{H})}$ on $\MW$ and $\sweff$ for various
MSSM parameter choices. The values are chosen such
that experimental constraints are fulfilled for most parts of the
parameter space.
\reffi{fig:LargeTB} shows a scenario with large $\tb$, 
$\tb=50$, and $\msusy = \MA = 300 \gev$ and $\mu=500\gev$. The
results are plotted as functions of the
stop-mixing parameter $\Xt=\At-\mu/\tb$ (see \refeq{stopmassmatrix}),
and we chose 
$\Ab=\At$. The two-loop contributions $\De\MW$ and $\De\sweff$ 
are decomposed into the SM result, 
$\De\rho^{\SM,\al_t^2}_{\rm 2-loop}(\MHSM = \Mh)$, 
as given in \refeq{drSMgf2} (shown with reversed sign
for better visibility),
$\De\rho^{(q)}({\rm MSSM-SM})$ as given in
\refeq{eq:delrhodiff}, and $\De\rho^{(\sq,\tilde{H})}$ as given in 
\refeq{massshiftincl}. For the latter contribution both the result
with the correct MSSM value for $\Mh$ and with $\Mh = 0$ is shown.
We find that $\De\rho^{(\sq,\tilde{H})}$ induces shifts in $\MW$ and
$\sweff$ of up to $+8 \mev$ in $\MW$ and $- 4 \times 10^{-5}$ in
$\sweff$. The corrections are significantly larger than the effective
change compared to the SM arising from class $(q)$, 
$\De\rho^{(q)}({\rm MSSM-SM})$. The impact of relaxing the gauge-less
limit on $\Mh$ in $\De\rho^{(\sq,\tilde{H})}$ is clearly visible,
although not as pronounced as in \reffi{fig:Mheq0}. 
It should be noted that small mixing in the stop sector (in this
scenario values of $|\Xt| \lsim 350 \gev$) is disfavored
by the LEP Higgs searches~\cite{LEPHiggsSM,LEPHiggsMSSM},
i.e.\ the dependence on $\Mh$
is largest where its value is already experimentally excluded.
For small values of $|\Xt|$ the supersymmetric contribution 
$\De\rho^{(q)}({\rm MSSM-SM}) + \De\rho^{(\sq,\tilde{H})}$ is
almost as large as the SM result,
$\De\rho^{\SM,\al_t^2}_{\rm 2-loop}(\MHSM = \Mh)$, 
and largely compensates it. For large
values of $|\Xt|$ the supersymmetric contribution
reduces the SM result by about 40\%.

In \reffi{fig:SmallMUE} we show a similar plot for a parameter scenario
with small Higgsino mass, $\mu=200\gev$, and $\tb = 6$,
$\msusy = 400 \gev$, $\MA = 300 \gev$. 
The contribution of $\De\rho^{(\sq,\tilde{H})}$ amounts to about
1--$2\mev$ in $\MW$ and $-1 \times 10^{-5}$ in $\sweff$ in this case.
The fermion loop contribution $\De\rho^{(q)}({\rm MSSM-SM})$
is very small here because the  
small value of $\tb$ does not lead to an enhancement of $\alb$ in the
MSSM with respect to the SM.

\begin{figure}[th!]
\begin{center}
\epsfig{figure=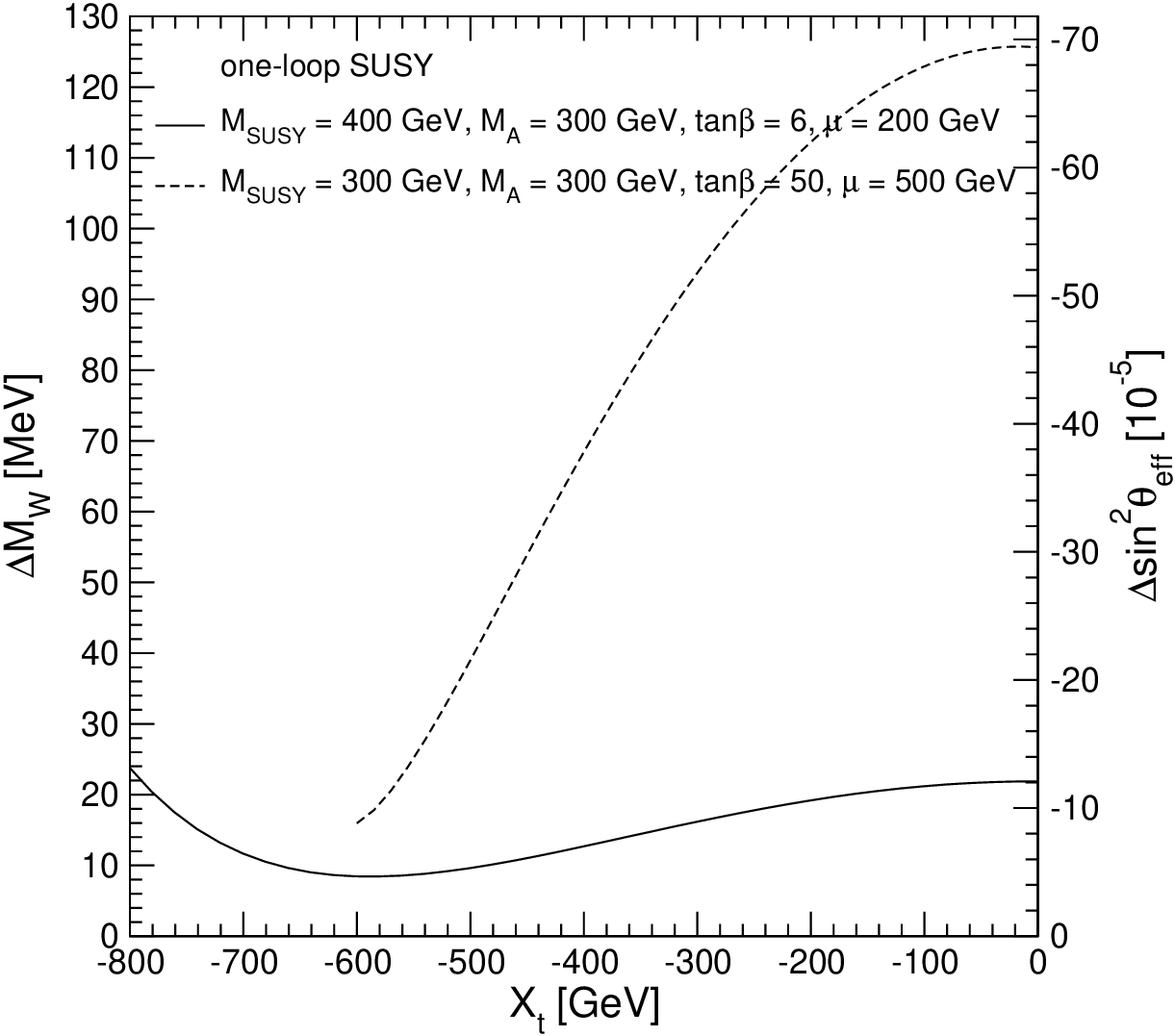,width=12cm,height=8.5cm}
\caption{%
One-loop SUSY contributions to $\De\MW$ and $\De\sweff$ are shown
as a function of $\Xt$. The parameters correspond to the two scenarios
analyzed in \reffis{fig:LargeTB} and \ref{fig:SmallMUE}.
}
\label{fig:oneloop}
\end{center}
\end{figure}

\reffi{fig:oneloop} 
shows the one-loop results, $\De\rho^{\rm SUSY}_{\rm 1-loop}$,
corresponding to the scenarios
of Figs.~\ref{fig:LargeTB}, \ref{fig:SmallMUE}. 
Due to the larger value of $\msusy$
and the small value of $\tb$ the one-loop contributions for the second
scenario are
relatively small. 
The region of small $|\Xt|$ is again ruled out by LEP Higgs searches. The
largest effects visible in \reffi{fig:oneloop} are thus experimentally
excluded. Comparing the one-loop with the two-loop results, one
can see that the two-loop contributions from 
$\De\rho^{(\sq,\tilde{H})}$ amounts to about
10\% of the one-loop supersymmetric contributions.

A common feature of the two scenarios, visible in Figs.\ \ref{fig:LargeTB},
\ref{fig:SmallMUE}, \ref{fig:oneloop}, 
is that both the one- and two-loop supersymmetric
contributions first decrease for increasing $|\Xt|$ until a
minimum is reached in the vicinity of $\Xt\sim-2\msusy$. For even
larger mixing one stop mass becomes very
small and the supersymmetric contributions increase again.


\subsection{Results in SPS scenarios and renormalization scheme\\
  dependence}

\begin{figure}[thb!]
\begin{center}
\epsfig{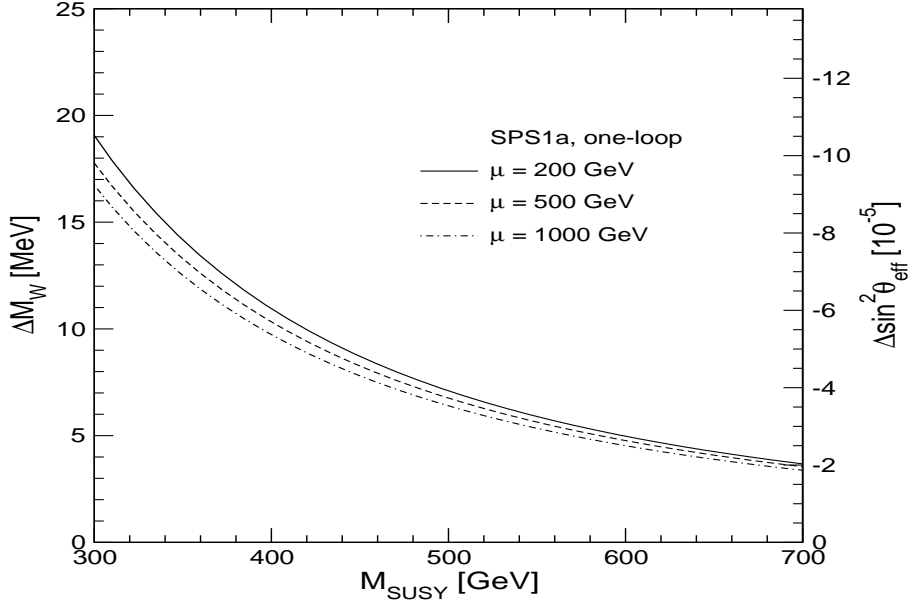}
\caption{%
The shifts $\De\MW$ and $\De\sweff$ induced by the supersymmetric one-loop 
contributions are shown as a
function of $\msusy$ in the 
SPS~1a scenario for $\mu = 200, 500, 1000 \gev$ and $\tb = 10$. 
}
\label{fig:oneloop2}
\end{center}
\end{figure}

\reffi{fig:Standard} shows the results 
for $\De\rho^{(\sq,\tilde{H})}$ in the SPS~1a
benchmark scenario~\cite{sps} 
for a moderate value of 
$\tb=10$ and four different combinations for $\mu$ and $\MA$, 
\BE
(\mu/{\rm GeV}, \MA/{\rm GeV}) = (200 , 200), (200 , 1000 ),
(500 , 500 ), (500 , 1000 ) . 
\label{eq:muMA}
\EE
In order to display the dependence on the scale of supersymmetry, 
we start from the
nominal values of the MSSM parameters corresponding to the SPS~1a
point~\cite{sps} (besides $\mu$ and $\MA$ that are chosen as specified
in \refeq{eq:muMA}) and vary the parameters
$\msusy$ and $A_{t,b}$ using a common scale factor; the results
are then shown as functions of $\msusy$.%
\footnote{
More precisely, for
  the SPS points the soft supersymmetry-breaking parameters
  $M_{\tilde{t}_L,\tilde{t}_R,\tilde{b}_R}$ for the left- and
  right-handed $\tilde{t}$, $\tilde{b}$ 
  are all slightly different. $\msusy$ is identified with
  $M_{\tilde{t}_L}$.
}
The range of $\msusy$ values shown in \reffi{fig:Standard} has been
chosen such that compatibility with
Higgs-boson mass~\cite{LEPHiggsSM} and $b$-physics~\cite{pdg}
constraints is ensured for most parts of the parameter space. 
For small values of $\msusy$ the corrections differ by up to $4 \mev$
depending on the choice of $\MA$ and $\mu$. Smaller values of $\MA$
and $\mu$ result in larger corrections to $\MW$ and $\sweff$. In all
cases the result decreases with increasing $\msusy$ as expected. 
The corresponding supersymmetric one-loop contributions 
induced by $\De\rho^{\rm SUSY}_{\rm 1-loop}$ are shown in
\reffi{fig:oneloop2} for comparison. The two-loop correction from 
$\De\rho^{(\sq,\tilde H)}$ amounts up to 25\% of the MSSM one-loop
result.

\begin{figure}[th!]
\begin{center}
\epsfig{figure=delrhoMSSM24A.bw.eps,width=12cm,height=8cm}
\caption{%
The shifts $\De\MW$ and $\De\sweff$ induced by $\De\rho^{(\sq,\tilde{H})}$
are shown as a function of $\msusy$ in the
SPS~1a scenario for four combinations of
and $\MA = 200, 500 \gev$ and $\mu = 200, 500, 1000 \gev$. 
}
\label{fig:Standard}
\end{center}
\end{figure}

\begin{figure}[htb!]
\begin{center}
\epsfig{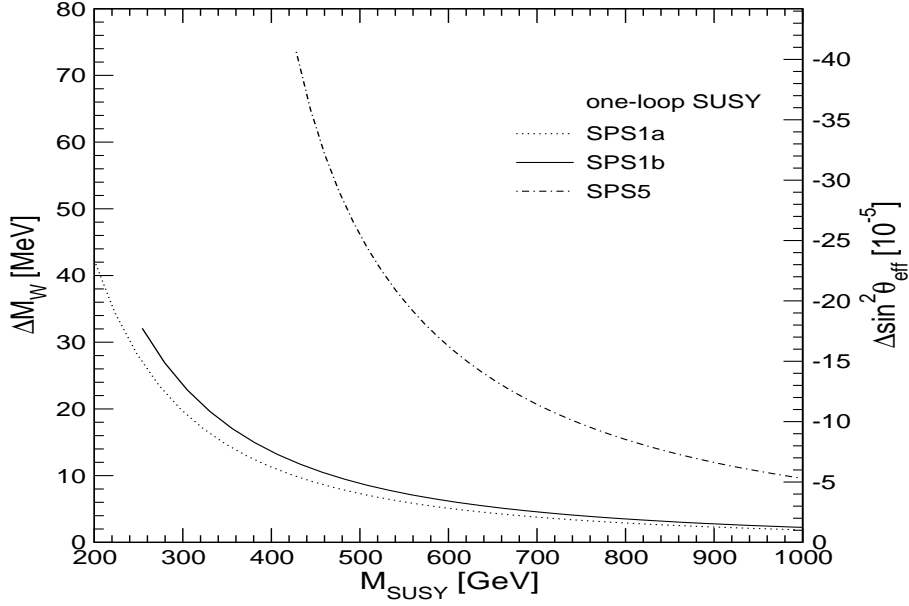}
\caption{%
The shifts $\De\MW$ and $\De\sweff$ induced by the supersymmetric one-loop 
contribution $\De\rho^{\rm SUSY}_{\rm 1-loop}$
are shown for the three benchmark
scenarios SPS~1a, SPS~1b and SPS~5 as a function of $\msusy$.
The parameters of the squark sector correspond to
the on-shell scheme. 
}
\label{fig:SPSoneloop}
\end{center}
\end{figure}

\begin{figure}[t!]
\begin{center}
\epsfig{figure=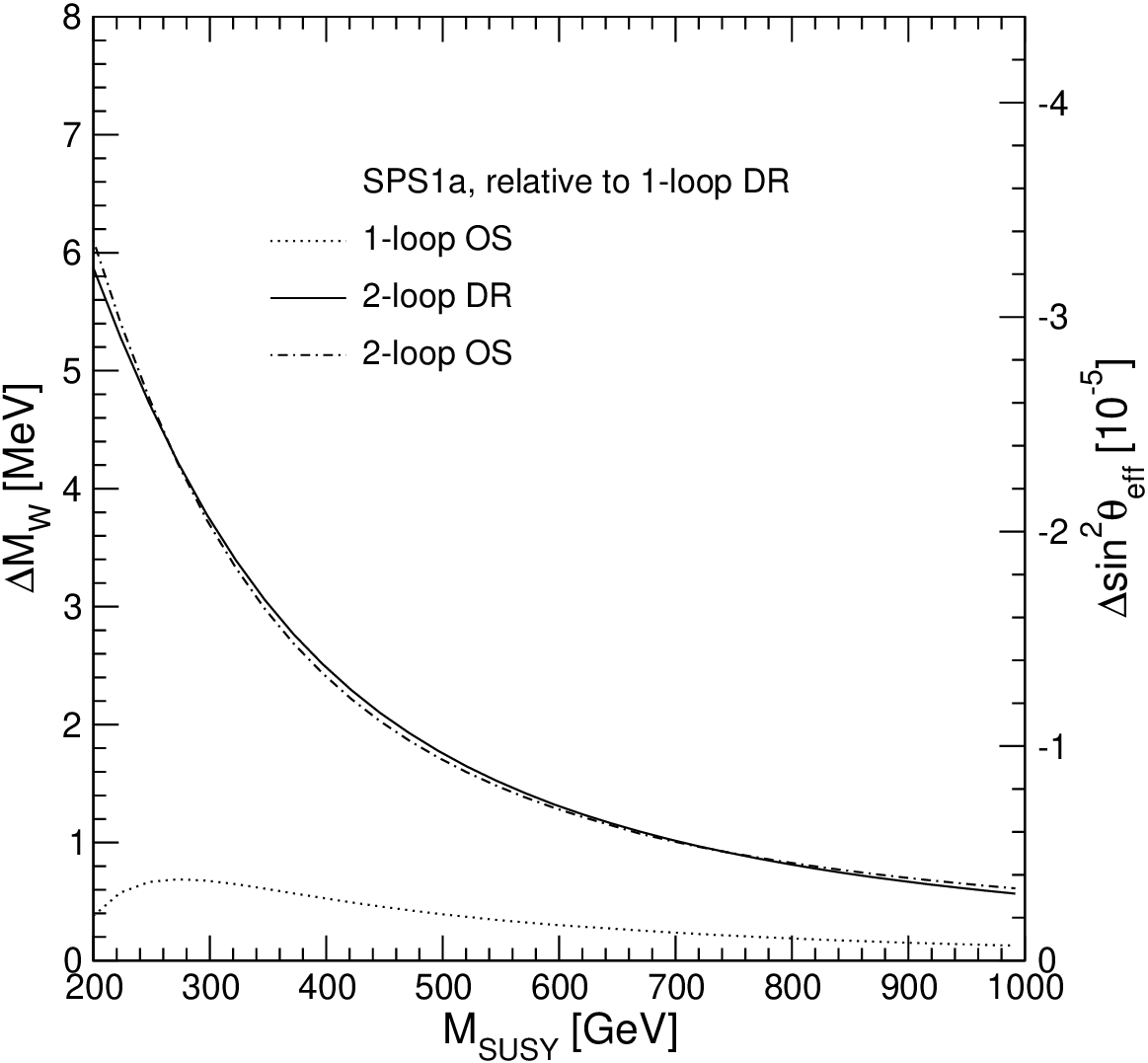,width=12cm,height=8cm}
\caption{%
$\De\MW$ and $\De\sweff$ are shown in the SPS~1a scenario as a function
of $\msusy$. The results for the one-loop contribution expressed in
terms of on-shell parameters and for the two-loop result 
$\De\rho^{\rm SUSY}_{\rm 1-loop}+\De\rho^{(\sq,\tilde{H})}$
in the on-shell and the \drbar\ scheme are given relative to the 
one-loop result expressed in terms of \drbar\ parameters, see
\refeq{SPSPlotsLines}.
}
\label{fig:SPS1a}
\end{center}
\end{figure}

\begin{figure}[th!]
\vspace{2em}
\begin{center}
\epsfig{figure=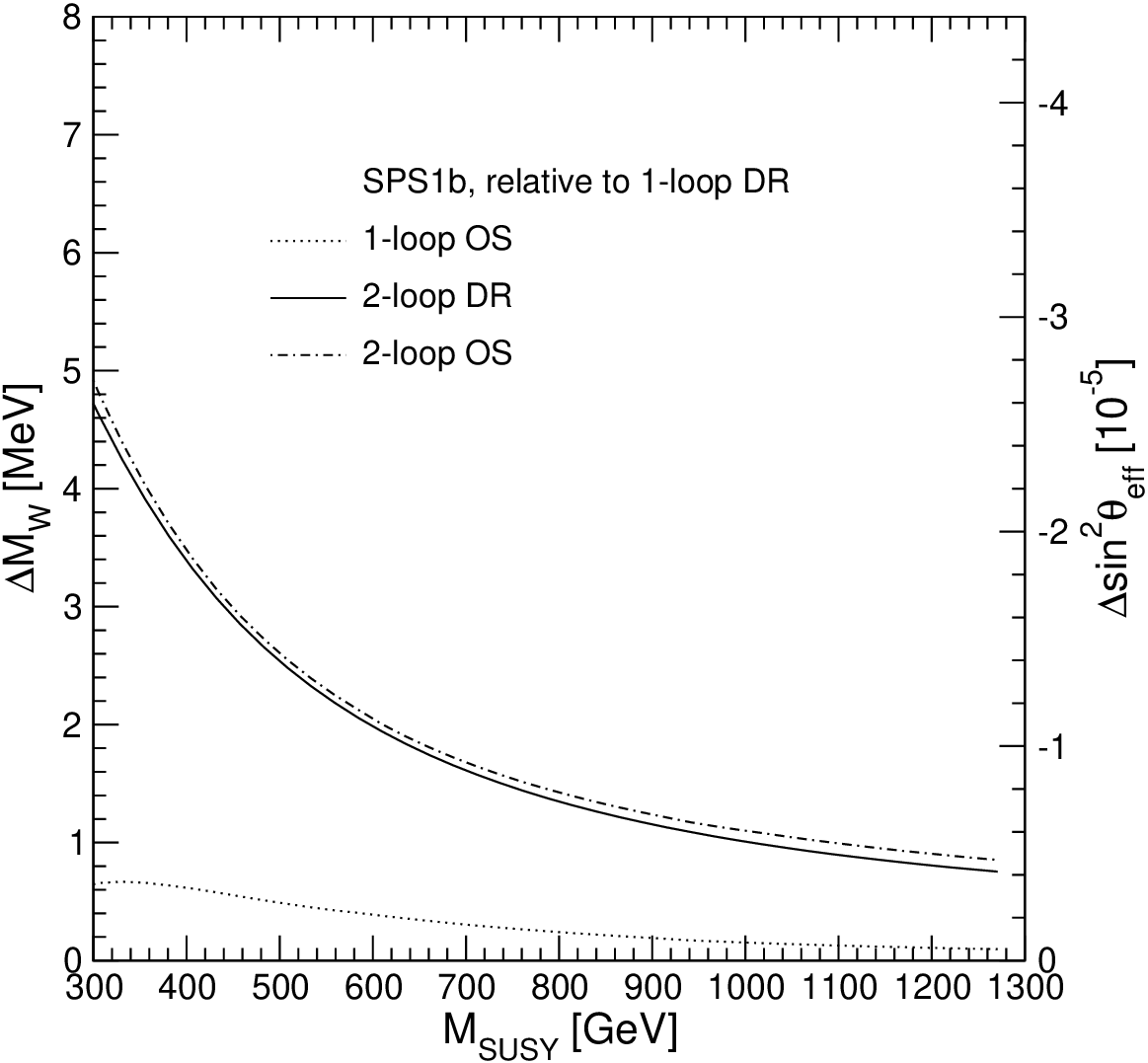,width=12cm,height=8cm}
\caption{%
$\De\MW$ and $\De\sweff$ are shown in the SPS~1b scenario as a function
of $\msusy$. The results for the one-loop contribution expressed in
terms of on-shell parameters and for the two-loop result 
$\De\rho^{\rm SUSY}_{\rm 1-loop}+\De\rho^{(\sq,\tilde{H})}$
in the on-shell and the \drbar\ scheme are given relative to the 
one-loop result expressed in terms of \drbar\ parameters, see
\refeq{SPSPlotsLines}.
}
\label{fig:SPS1b}
\end{center}
\end{figure}

\begin{figure}[th!]
\begin{center}
\epsfig{figure=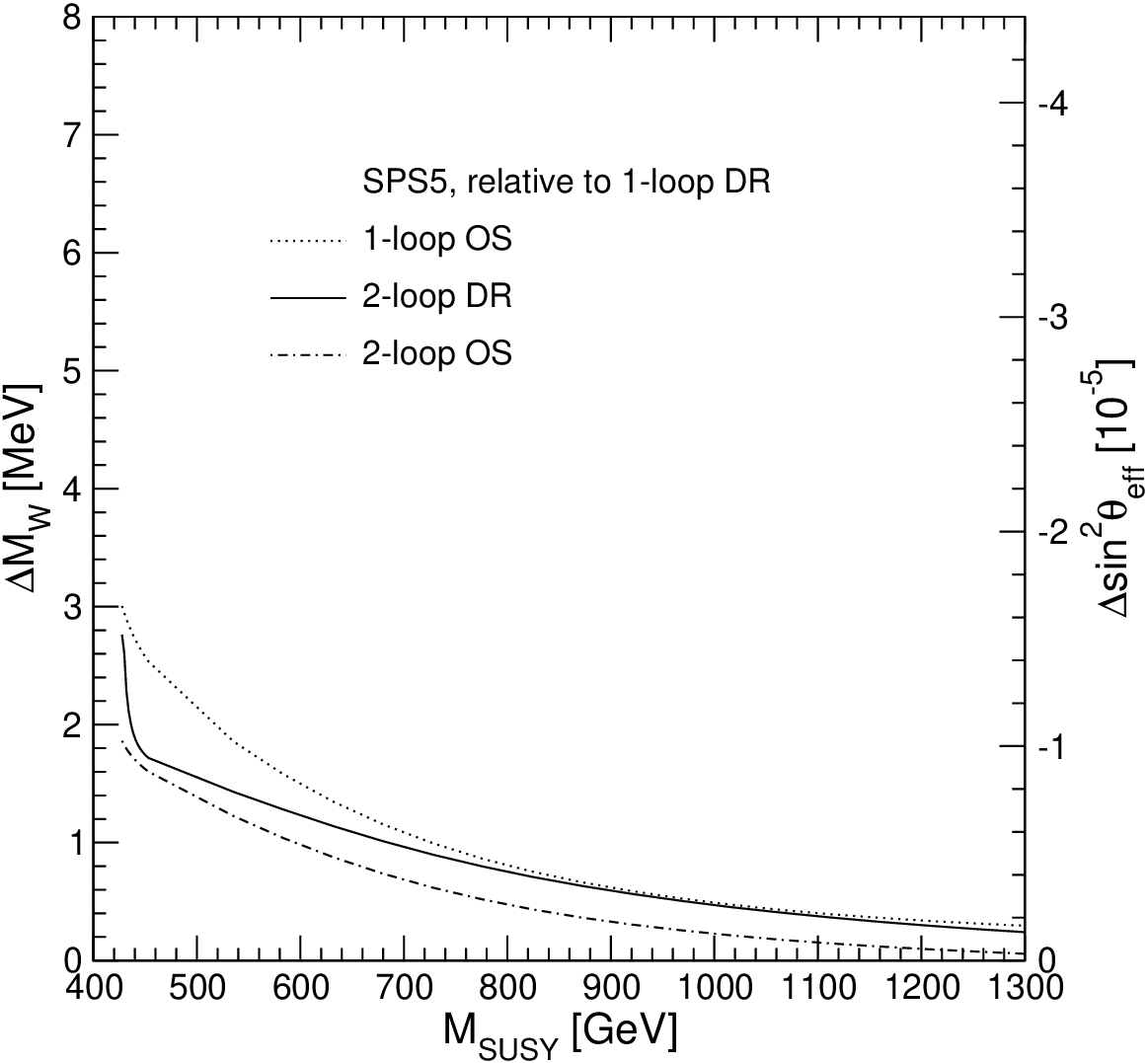,width=12cm,height=8cm}
\vspace{2em}
\caption{%
$\De\MW$ and $\De\sweff$ are shown in the SPS~5 scenario as a function
of $\msusy$. The results for the one-loop contribution expressed in
terms of on-shell parameters and for the two-loop result 
$\De\rho^{\rm SUSY}_{\rm 1-loop}+\De\rho^{(\sq,\tilde{H})}$
in the on-shell and the \drbar\ scheme are given relative to the 
one-loop result expressed in terms of \drbar\ parameters, see
\refeq{SPSPlotsLines}.
}
\label{fig:SPS5}
\end{center}
\end{figure}

We now study the renormalization scheme dependence of the one-loop and
two-loop results for three benchmark SPS scenarios. Besides the ``standard''
scenario SPS~1a, we also investigate the SPS~1b scenario, which is
characterized by a larger $\tb$ value, $\tb=30$, and SPS~5, which
involves a relatively light $\Stop$~\cite{sps}. \reffi{fig:SPSoneloop}
shows the one-loop results for the three scenarios, while
\reffis{fig:SPS1a}, \ref{fig:SPS1b}, \ref{fig:SPS5} display the two-loop
results. As above, the results are shown as functions of $\msusy$. We
have started from the nominal values of the MSSM parameters for the
three benchmark points and varied the parameters $\msusy$, $A_{t,b}$,
$\mu$ (for the \drbar\ results also the scale $\mu^{\drbarm}$)
using a common scale factor. 
The actual SPS~1a, SPS~1b and SPS~5 benchmark points correspond to 
$\msusy = 495.9$, 762.5, 535.2$\gev$, respectively~\cite{sps}.

For a meaningful comparison of the results in the on-shell and the
\drbar\ renormalization schemes, the input
parameters in the two schemes have to be physically equivalent, which
implies that they are numerically different. Since the parameters
in the SPS scenarios are defined as \drbar\ parameters, they can
directly be used as input parameters in the \drbar\ scheme.
The corresponding input parameters for the calculation in
the on-shell scheme are obtained by requiring 
\BE
(m_{\tilde{f}_i}^2+\de
m_{\tilde{f}_i}^2)^{\rm OS}=(m_{\tilde{f}_i}^2+\de
m_{\tilde{f}_i}^2)^{\rm \overline{{\rm DR}}}
\EE
for the squark masses and similarly for
the mixing angles.

In the one-loop results $\De\rho^{\rm SUSY}_{\rm 1-loop}$ for the three 
SPS scenarios shown in \reffi{fig:SPSoneloop}
the squark sector parameters correspond to the on-shell scheme. 
The shift in the precision observables induced by 
$\De\rho^{\rm SUSY}_{\rm 1-loop}$
is found to be particularly large for the SPS~5 scenario, as a
consequence of the large splitting between the squark masses in this
scenario.

In \reffis{fig:SPS1a}--\ref{fig:SPS5} 
we show the one-loop result parametrized in terms of
on-shell parameters (dotted line)
and the two-loop $(\sq,\tilde{H})$ results obtained in the \drbar\ 
(full line) and the OS~scheme (dot-dashed line), in all cases relative to
the one-loop result parametrized in terms of the \drbar\ parameters. 
Accordingly, the three lines in each plot correspond to
\BE
\left\{
\begin{array}{l}
\De\rho^{\rm SUSY,OS}_{\rm 1-loop} \\[.5ex]
\De\rho^{\rm SUSY,\drbarm}_{\rm 1-loop} +
   \De\rho^{(\sq,\tilde{H}),\drbarm}  \\[.5ex]
\De\rho^{\rm SUSY,OS}_{\rm 1-loop} +
   \De\rho^{(\sq,\tilde{H}),{\rm OS}}
\end{array}
\right\}
 -\De\rho^{\rm SUSY,\drbarm}_{\rm 1-loop}.
\label{SPSPlotsLines}
\EE
The pure two-loop correction in the \drbar\ scheme is given by the full
line, while the two-loop correction in the on-shell scheme corresponds
to the difference between the dot-dashed and the dashed line.

The numerical impact of the two-loop correction
$\De\rho^{(\sq,\tilde{H})}$ in the scenarios SPS~1a,~1b amounts to about
5--$6 \mev$ in $\MW$ and $-3 \times 10^{-5}$ in $\sweff$ for small
$\msusy$ and decreases to about $1 \mev$ in $\MW$ ($-0.5 \times 10^{-5}$
in $\sweff$) for larger values of $\msusy$.
For SPS~5 the corrections are slightly smaller. 
While in the scenarios SPS~1a,~1b the two-loop results in the two
schemes are very close to each other, a larger deviation is visible in
the SPS~5 scenario. In the latter scenario
the two-loop corrections in the
on-shell scheme are less than $1\mev$, while in the \drbar\ scheme
they are more than twice as large.
Comparison with the one-loop results given in
\reffi{fig:SPSoneloop} shows that the two-loop corrections amount to
about 10\% one-loop MSSM
contribution.

The comparison of the renormalization schemes shows that the scheme
dependence is strongly reduced by going from the one-loop to the
two-loop level. At the one-loop level, where the scheme difference 
is entirely due to the different input parameters for
the squark masses and mixing angles, the difference between the on-shell
and the \drbar\ scheme is of \order{1\mev} in $\MW$. Taking into account the
two-loop corrections reduces the difference below $0.1\mev$ for SPS~1a,b
and about $0.2\mev$ for SPS~5.

The size of the two-loop corrections for SPS~1a,b is found to be much larger 
than the difference between the two schemes at the one-loop level, 
which is only about $1 \mev$ for these scenarios. This indicates that
the difference between the results in two renormalization schemes,
if taken as the only measure for estimating the theoretical uncertainties from
unknown higher-order corrections, may result in a significant
underestimate of the actual theoretical uncertainty.
The SPS~5 scenario, on the other hand, is an example where 
the two-loop corrections turn out to be smaller than the scheme
difference at one-loop order.

\begin{figure}[th!]
\begin{center}
\epsfig{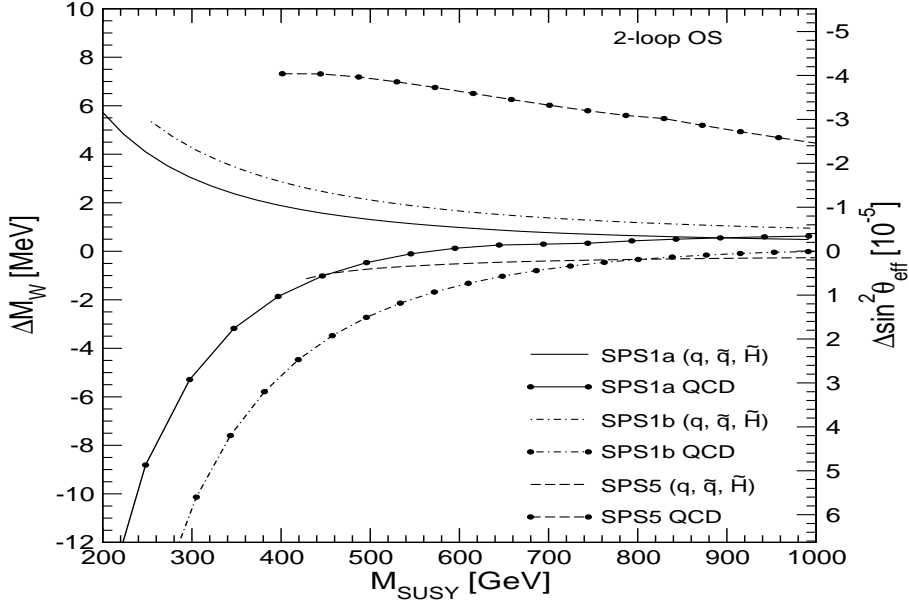}
\caption{%
The effect of the two-loop Yukawa corrections from squark and higgsino
loops is compared with the squark-loop corrections of \order{\al\als}.
$\De\MW$ and $\De\sweff$ are shown in the three
SPS scenarios as a function of $\msusy$ in the on-shell scheme.
}
\label{fig:QCDcomp}
\end{center}
\end{figure}

Finally we compare the two-loop results 
for the $(\sq,\tilde{H})$ contributions obtained in this paper with
the two-loop QCD corrections of \order{\al\als} as obtained in
\citere{dr2lA}. In \reffi{fig:QCDcomp} we show the results in
the on-shell scheme for the three SPS scenarios as a function of
$\msusy$ (as explained above). For SPS~1a and~1b both corrections are
roughly of the same size and compensate each other to a large extent. 
Only for the case of SPS~5 the QCD corrections are
significantly larger than the two-loop Yukawa corrections. Both the QCD
and the Yukawa corrections are non-negligible in view of the anticipated
future experimental accuracies and larger than the current theoretical
uncertainties in the SM.

\subsection{Estimate of unknown higher-order corrections}
\label{subsec:intrinsicfuture}

As discussed above, the theoretical evaluation of the EWPO in the SM is
significantly more advanced than in the MSSM. In order to obtain an
accurate prediction for the EWPO within the MSSM it is therefore useful to 
take all known SM corrections into account. This can be done by writing
the MSSM prediction for the observable $O$ ($O = \MW, \sweff, \ldots$) 
as
\BE
O_{\rm MSSM} = O_{\rm SM} + O_{{\rm MSSM}-{\rm SM}}~,
\label{eq:obsSMSUSY}
\EE
where $O_{\rm SM}$ is the prediction in the SM including all known
corrections, and $O_{{\rm MSSM}-{\rm SM}}$ is the difference
between the MSSM and the SM predictions, evaluated at the level of
precision of the known MSSM corrections. The expression given in
\refeq{eq:obsSMSUSY} contains higher-order contributions that are only
known for SM particles in the loop but not for their superpartners
(e.g.\ two-loop electroweak corrections beyond the leading Yukawa
contributions calculated in this paper and three-loop corrections of
\order{\al\als^2}). In the decoupling limit where all superpartners
are heavy and the Higgs sector becomes SM-like, the result of 
\refeq{eq:obsSMSUSY} obviously yields a more precise prediction than a 
result based on only those corrections which are known in the full MSSM. 
In this case the second term in \refeq{eq:obsSMSUSY} goes to zero, so
that the MSSM result approaches the SM result with $\MHSM = \Mh$.
For lower values of the scale of supersymmetry the contribution from 
supersymmetric particles in the the loop may be of comparable size as
the known SM corrections. In view of the experimental bounds on the
masses of the supersymmetric particles (and the fact that supersymmetry
has to be broken), however, a complete cancellation between
the SM and supersymmetric contributions is not expected. It therefore
seems appropriate to apply \refeq{eq:obsSMSUSY} also in this case.

Expressing the predictions for the EWPO as in \refeq{eq:obsSMSUSY}
implies that the theoretical uncertainties from unknown higher-order
corrections reduce to those in the SM in the decoupling limit. In the
SM, based on all higher-order contributions that are currently known,
the remaining uncertainties in $\MW$~\cite{MWSM} and
$\sweff$~\cite{sweffSM} have been estimated to be
\BE
\de\MW^{\rm SM} = 4 \mev, \quad \de\sweff^{\rm SM} = 5 \times 10^{-5}~.
\label{eq:SMunc}
\EE
Below the decoupling limit an additional
theoretical uncertainty arises from higher-order corrections involving
supersymmetric particles in the loops. In the following we will estimate 
this additional theoretical uncertainty in the prediction of $\MW$ and
$\sweff$ depending on
the supersymmetric parameters. We will provide estimates for the
uncertainty for three
values of the squark mass scale, $\msusy = 200$, 500, $1000 \gev$.
A similar approach of estimating the remaining uncertainties from
unknown higher-order corrections with dependence on the supersymmetric
parameters has recently been applied to the Higgs sector and implemented
in the program \fh 2.2, see \citere{feynhiggs2.2} for details.

The remaining uncertainties from unknown higher-order corrections
involving supersymmetric particles mainly arise from the following
sources:

\begin{itemize}

\item
Electroweak two-loop corrections beyond the leading Yukawa corrections
evaluated in this paper:\\
We estimate the numerical effect of these corrections by assuming that
the ratio of the subleading electroweak two-loop corrections to the
two-loop Yukawa corrections is the same in the SM as in the MSSM.
Inserting the known SM corrections~\cite{deltarSM,MWSM} we infer an 
estimate of the possible size of the missing supersymmetric electroweak 
two-loop contributions.

\item
\order{\al\als} corrections beyond the $\De\rho$ approximation:\\
We estimate the size of these corrections by assuming that the ratio of
the contribution entering via $\De\rho$ to the full result is the same
as for the known SM result~\cite{deltarSMgfals}.

\item
\order{\al\als^2} corrections:\\
We use three different methods for estimating the possible size of these 
corrections. The unknown ratio of the \order{\al\als^2} supersymmetric
contributions to the \order{\al\als} supersymmetric contributions can be
estimated by assuming that it is the same as for the corresponding
corrections in the SM~\cite{drSMgfals2} (estimate $(a)$)
and, using geometric progression
from lower orders, by assuming that it is the same as the ratio of the
\order{\al\als} supersymmetric contributions and the \order{\al}
supersymmetric contributions (estimate $(b)$). 
As a further indication of the possible
size of unknown corrections of \order{\al\als^2} we vary
the renormalization scale of $\als(\mu^{\drbarm})$ entering the
\order{\al\als} result according to $\mt/2 \leq \mu^{\drbarm} \leq 2 \mt$
(estimate $(c)$).
It should be noted that this variation of $\als(\mu^{\drbarm})$ corresponds
to only a part of the higher-order corrections, so that estimates $(a)$
and $(b)$ should
be regarded as more conservative.

\item
\order{\al^2\als} corrections:\\
Similarly as for the \order{\al\als^2} corrections, we again use three
different methods for estimating these corrections. The unknown ratio of
the \order{\al^2\als} supersymmetric contributions to the \order{\al^2} 
(leading Yukawa) supersymmetric contributions can be
estimated by assuming that it is the same as for the corresponding
corrections in the SM~\cite{drSMgf3} 
(estimate $(a)$)
and by assuming that it is the same as the ratio of the
\order{\al\als} supersymmetric contributions and the \order{\al}
supersymmetric contributions
(estimate $(b)$). As a further indication of possible
corrections of \order{\al^2\als} we change the value of $\mt$ in the 
result for the two-loop supersymmetric Yukawa corrections from the 
on-shell value, $\mt^{\rm OS}$, to the running mass $\mt(\mt)$, where
$\mt(\mt) = \mt^{\rm OS}/(1 + 4/(3 \pi) \, \als(\mt))$ (estimate $(c)$).
The latter replacement accounts only for a subset of the unknown
\order{\al^2\als} corrections.

\item
Electroweak three-loop corrections:\\
As an indication of the possible size of these corrections we use 
the renormalization
scheme dependence of our result for the supersymmetric two-loop Yukawa
corrections, see \reffis{fig:SPS1a}--\ref{fig:SPS5}. 

\end{itemize}

We have evaluated the above estimates for the three scenarios SPS~1a,
SPS~1b, and SPS~5, each for $\msusy = 1000\gev$, $500\gev$, and for
$\msusy<500\gev$~%
\footnote{%
The lowest values considered for $\msusy$ are $200$, 300, 400$\gev$
for SPS1a, SPS1b, SPS5, respectively. These are the lowest values
shown in Figs.\ \ref{fig:SPS1a}, \ref{fig:SPS1b}, \ref{fig:SPS5}.
For lower values the parameter points are 
excluded by Higgs mass constraints.}
~(as above we have varied $\msusy$, $A_{t,b}$ and $\mu$ using a common
scale factor). The estimated theoretical uncertainties for $\MW$ arising
from the different classes of unknown higher-order corrections are shown
in \refta{tab:MWunc}. The result given in each entry
corresponds to the largest value obtained in the three considered SPS
scenarios. The three numbers given for the \order{\al\als^2} and 
\order{\al^2\als} corrections correspond to the estimates $(a)$, $(b)$ and
$(c)$ described above.

\begin{table}[tbh!]
\renewcommand{\arraystretch}{1.5}
\BC
\begin{tabular}{|c|c|c|c|}
\hline
$\msusy$ &  $<$500 GeV  &  500 GeV  &  1000 GeV  \\
\hline\hline
\order{\al^2} subleading   & 6.0 & 2.0 & 0.8 \\ \hline
\order{\al\als} subleading & 1.8 & 0.9 & 0.5 \\ \hline
\order{\al\als^2}   & 3.0, \; 5.3, \; 1.5 & 1.4, \; 1.1, \; 0.7 &
                      0.9, \; 2.2, \; 0.5 \\ \hline
\order{\al^2\als}   & 1.5, \; 2.2, \; 1.4 & 0.6, \; 0.8, \; 0.4 &
                      0.2, \; 0.2, \; 0.2 \\ \hline
\order{\al^3}              & 0.3 & 0.3 & 0.3 \\ 
\hline\hline
\end{tabular}
\EC
\renewcommand{\arraystretch}{1}
\caption{
Estimated uncertainties for $\MW$ in MeV for different classes of
unknown higher-order corrections involving supersymmetric particles are
given for three values of
$\msusy$. The estimates have been obtained using the results for the 
SPS~1a, SPS~1b, and SPS~5 scenarios. The three entries for the 
\order{\al\als^2} and \order{\al^2\als} corrections 
correspond to three different
methods for estimating the uncertainties (see text).
}
\label{tab:MWunc}
\end{table}

As expected, the estimated uncertainties associated with the
supersymmetric higher-order contributions decrease for increasing
$\msusy$. For the \order{\al\als^2} and \order{\al^2\als} corrections, 
method $(c)$ that accounts only for a part of the higher-order
corrections yields in both cases the most optimistic estimate.
As discussed earlier, by taking into account the true MSSM-value
of $\Mh$, certain parts of the electroweak corrections,
beyond the leading two-loop Yukawa corrections, are included
in our result. The difference between 
$\De\rho^{(\sq,\tilde{H})}(\Mh)$ and $\De\rho^{(\sq,\tilde{H})}(0)$ may
be interpreted as an estimate of the size of further, not included
higher-order electroweak corrections. The numerical analysis
in \refses{sec:51} and \ref{subsec:ewponum} shows that this estimate
is typically smaller than the estimated total uncertainty in
\refta{tab:MWunc}.

We now combine the values given in \refta{tab:MWunc} into our total
estimate of the remaining theoretical uncertainties from unknown 
higher-order corrections involving supersymmetric particles.
Adopting the largest of the three values for the \order{\al\als^2} and
\pagebreak[3]
\order{\al^2\als} as a conservative error estimate and adding the
different estimates in quadrature we obtain
\BEA
\de\MW &=& 8.5 \mev \mbox{ for } \msusy < 500 \gev , \non \\
\de\MW &=& 2.7 \mev \mbox{ for } \msusy = 500 \gev ,  \label{eq:MWunc}\\
\de\MW &=& 2.4 \mev \mbox{ for } \msusy = 1000 \gev . \non 
\EEA

An analogous analysis of the remaining higher-order uncertainties can
also be carried out for $\sweff$. Since parts of the missing
higher-order corrections to $\sweff$ and $\MW$ are related to each
other, we employ \refeq{precobs1} to infer estimates for $\sweff$ from
our results for $\MW$. This yields
\BEA
\de\sweff &=& 4.7 \times 10^{-5} \mbox{ for } \msusy < 500 \gev , \non \\
\de\sweff &=& 1.5 \times 10^{-5} \mbox{ for } \msusy = 500 \gev ,
\label{eq:sweffunc} \\
\de\sweff &=& 1.3 \times 10^{-5} \mbox{ for } \msusy = 1000 \gev . \non
\EEA

The full theory uncertainty in the MSSM can be obtained by adding in
quadrature the SM uncertainties from eq.\ (\ref{eq:SMunc}) and the
SUSY uncertainties from eqs.\
(\ref{eq:MWunc})--(\ref{eq:sweffunc}). This yields $\delta
\MW=(4.7\,-\,9.4)\mev$ and $\delta\sweff=(5.2\,-\,6.7)\times10^{-5}$
depending on the SUSY mass scale.

The estimated uncertainties 
are smaller than the estimates in \citere{PomssmRep} (where an overall
estimate has been given without analyzing the dependence on the
supersymmetric parameters), reflecting the improvement associated with
the new corrections calculated in this paper.

The other source of theoretical uncertainties besides the one from
unknown higher-order corrections is the parametric uncertainty induced
by the experimental errors of the input parameters. The current
experimental error of the top-quark mass~\cite{mtexpnew} induces the
following parametric uncertainties in $\MW$ and $\sweff$
\BE
\de\mt^{\rm exp} = 2.9 \gev \;\; \Rightarrow \;\; 
\de\MW^{{\rm para}, \mt} = 17.5 \mev,  \;\;
\de\sweff^{{\rm para}, \mt} = 9.4 \times 10^{-5}~.
\EE
This uncertainty will decrease during the next years as a consequence of 
a further improvement of the accuracy on $\mt$ at the Tevatron and the
LHC. Ultimately it will be reduced by more than an order of magnitude at
the ILC~\cite{deltamt}. The accuracy of the theoretical predictions for
$\MW$ and $\sweff$ will then be limited by the uncertainty from unknown
higher-order corrections (for a discussion of the parametric uncertainties 
induced by the other SM input parameters see \citere{PomssmRep}). A
further reduction of the uncertainties from higher-order SM-type
corrections (see \refeq{eq:SMunc}) and corrections involving
supersymmetric particles (see \refeqs{eq:MWunc}--(\ref{eq:sweffunc}))
therefore seems to be in order to fully exploit the prospective
experimental accuracies on $\MW$, $\sweff$ and $\mt$ reachable at the
next generation of colliders~\cite{deltamt,gigaz}.


\section{Conclusions}
\label{sec:conclusions}

In this paper we have calculated the two-loop corrections of 
\order{\alt^2}, 
\order{\alt \alb}, \order{\alb^2} to the electroweak precision
observables $\MW$ and $\sweff$ in the MSSM. These are the leading,
Yukawa-enhanced 
electroweak two-loop contributions; they enter via $\De\rho$ and arise
from diagrams involving SM quarks, squarks, Higgs bosons and higgsinos.
While previously only the contribution from
the diagrams with quarks and Higgs bosons had been known (corresponding
to the limiting case where all supersymmetric particles are infinitely
heavy), we have evaluated the complete set of Yukawa corrections
including the effects of supersymmetric particles.

We have given a detailed account of the theoretical basis
of the calculation, focusing on the implications of the parameter
relations enforced by supersymmetry. In the gauge-less limit that needs
to be employed to extract the Yukawa corrections of \order{\alt^2},
\order{\alt \alb}, \order{\alb^2} the 
lightest MSSM Higgs boson mass $\Mh$ vanishes. We have studied in how
far the true MSSM value for $\Mh$ can be taken into account in a
consistent way. We have shown that the result can be expressed in such a
way that the $\Mh$-dependence, being formally a sub-leading effect, can
be kept essentially everywhere and we have compared this result with the
case where the gauge-less limit is sctrictly imposed. 

In our numerical analysis we have put the main emphasis on the new 
supersymmetric contributions involving squarks and higgsinos. We
have analyzed the results of the new contributions as functions of the
squark mass scale $\msusy$, the stop mixing $\Xt$ and the higgsino and
Higgs boson mass parameters $\mu$ and $\MA$. For squark masses of
about $300\gev$ we find corrections of typically $+4 \mev$ in $\MW$ and
$-2 \times 10^{-5}$ in $\sweff$. In certain parameter regions, in
particular slightly smaller values of $\msusy$ or small mixing in the
stop sector, we find shifts up to $+8 \mev$ in $\MW$ and 
$-4 \times 10^{-5}$ in $\sweff$. 
For a wide range of parameters, the squark and higgsino
two-loop corrections   increase  the corresponding one-loop 
contributions by about 10\%. 

Therefore, the class of diagrams with squarks and higgsinos, which has
no SM counterpart,  gives rise to significant deviations from
the SM predictions. This is in contrast with the contribution of the
diagrams involving quarks and Higgs bosons, which can be well
approximated by the corresponding SM contribution (setting the SM
Higgs-boson mass equal to the mass of the lightest $\cp$-even Higgs
boson of the MSSM). We have compared our result for the two-loop
Yukawa correction of \order{\alt^2}, \order{\alt \alb}, \order{\alb^2}
to $\MW$ and $\sweff$ with the \order{\al\als} 
correction, which is the only other genuine two-loop contribition to
$\MW$ and $\sweff$ known in the full MSSM. We find that the two
corrections are of comparable size and can largely compensate each
other for small values of $\msusy$ (depending on the other
supersymmetric parameters). 

We have derived our result in two renormalization schemes, the
on-shell scheme and the \drbar\ scheme for the squark sector
parameters. Comparing the two-loop results with the one-loop result
expressed in terms of the parameters of the two schemes shows a 
significant reduction of the scheme dependence. 

We have shown how the known corrections to the electroweak precision
observables in the SM and the MSSM can be combined such that the
currently most accurate prediction in the MSSM is obtained. In the
decoupling limit, where all supersymmetric particles are heavy, the 
theoretical uncertainty from unknown higher-order corrections reduces to
the uncertainty of the SM contribution. For non-vanishing contributions
of the supersymmetric particles an additional theoretical uncertainty
arises from unknown higher-order corrections involving supersymmetric 
particles.

We have estimated the current uncertainty from unknown higher-order
corrections involving supersymmetric particles for different values of the
squark mass scale $\msusy$. This has been done using geometric
progression from lower orders, employing known results for corresponding
SM corrections, investigating the renormalization scheme dependence, 
varying the renormalization scale, and taking into account formally
subleading $\Mh$-dependent contributions. For a squark mass scale below
$500 \gev$ we obtain an estimated uncertainty of about
$8.5 \mev$ in $\MW$ and $4.5 \times 10^{-5}$ in $\sweff$. These
uncertainties reduce to about $2.5 \mev$ in $\MW$ and $1.5 \times 10^{-5}$
in $\sweff$ for $\msusy = 1 \tev$. They can be combined quadratically
with the theory uncertainty from unknown higher-order SM contributions
to obtain the full MSSM theory uncertainties. While currently these
uncertainties  
(for $\msusy < 500 \gev$) are about a factor of two smaller than the 
parametric theoretical uncertainties induced by the experimental error
of the top-quark mass, their impact will become more pronounced with the
expected improvement of the experimental precision of $\mt$. 
The new two-loop corrections evaluated in this paper have been important
to reduce the theoretical uncertainties to the present level. Further
efforts on higher-order corrections in the MSSM will be necessary in
order to reduce the theoretical
uncertainties from unknown higher order corrections within the MSSM to
the level that has been reached for the SM.


\subsection*{Acknowledgements}

We thank W.~Hollik and A.~Weber for helpful discussions. S.H.\ and G.W.\
thank the Max Planck Institut f\"ur Physik, M\"unchen, for kind hospitality 
during part of this work.


\end{document}